\documentclass[a4paper,11pt]{article}
\pdfoutput=1
\usepackage{jcappub}

\usepackage{xcolor}
\definecolor{revisionblue}{RGB}{0,80,180}
\newcommand{\black}[1]{{\color{black}#1}}
\newenvironment{bluechange}{\color{black}}{}

\DeclareSymbolFont{myletters}{OML}{ztmcm}{m}{it}
\DeclareMathSymbol{\uplambda}{\mathord}{myletters}{"15}
\DeclareMathSymbol{\upxi}{\mathord}{myletters}{"18}

\renewcommand{\figurename}{FIGURE}

\usepackage{lmodern} 
\usepackage{booktabs}
\definecolor{Gray}{gray}{0.9}
\usepackage[includemp,
paperwidth=21.5cm,
paperheight=29.7cm,
top=2.30cm,
bottom=2.3cm,
inner=3cm,
outer=2.0cm,
marginparwidth=0.3cm,
marginparsep=0.3cm]{geometry}

\usepackage{setspace}
\definecolor{revisionred}{RGB}{0,0,0}

\newenvironment{revision}{\color{revisionred}}{}

\begin{document}
\title{Entanglement, Discord, and Residual Coherence in Scalar-Induced Gravitational Waves}

\author[]{Waqas Ahmed $^{\dagger}$}

\affiliation[a]{\it Center for Fundamental Physics and School of Artificial Intelligence, Hubei Polytechnic University, Huangshi 435003, China}

\emailAdd{$^\dagger$waqasmit@hbpu.edu.cn}


\abstract{
Scalar-induced gravitational waves (SIGWs) are usually characterized by their power spectrum, although their quadratic scalar source also carries higher-order correlation information. We investigate whether phase-sensitive correlations of a primordial Gaussian scalar state can survive decoherence and be transferred to the induced tensor sector. The scalar state is described in a fixed oscillator basis by the occupation $N_k^\zeta$ and anomalous moment $M_k^\zeta$, while the transfer is formulated through the unequal-time scalar Wightman function. The late-time tensor moments $d_k$ and $\gamma_k$ define an effective Gaussian reference state and its discord, whereas the exact SIGW state is generally non-Gaussian and contains an additional connected fourth-order cumulant. We therefore introduce the baseline-subtracted observable
$\Delta\kappa_{\rm coh}=\kappa_N[G^>_{\rm coh}]-\kappa_N[G^>_{\rm ref}]$.
For phase damping with $M_k^\zeta\propto e^{-D_k}$, the Gaussian-reference contribution conditionally scales as
$|\gamma_k|\propto e^{-2D_k}$ and $\kappa_N^{\rm G}\propto e^{-4D_k}$.
This provides a framework in which the SIGW spectrum fixes the signal band, while higher-order statistics probe residual coherence beyond the power spectrum.

}

\maketitle

\section{Introduction}
\label{sec:introduction}

One of the most striking lessons of inflation is that the observed structure of
the Universe may have begun as quantum fluctuations. In the standard picture,
vacuum fluctuations of the curvature perturbation are stretched to
super-Hubble scales, amplified, and later re-enter the horizon as the seeds of
cosmic microwave background anisotropies and large-scale structure
\cite{Guth:1980zm,Linde:1981mu,Albrecht:1982wi,Mukhanov:1981xt,Mukhanov:1990me,Polarski:1995jg}.
This gives inflation a remarkable status: it connects microscopic quantum
physics with macroscopic cosmological observables. Yet the observational
meaning of this statement is still subtle. By the time primordial perturbations
are measured, they are usually described extremely well by classical stochastic
variables. Whether any genuinely quantum information can survive this
transition and appear in late-time data remains an open and important question.

The usual discussion focuses on entanglement between opposite Fourier modes.
Such entanglement is naturally generated during inflation because the
perturbations are produced in highly squeezed two-mode states
\cite{Grishchuk:1990bj,Albrecht:1992kf,Polarski:1995jg,Kiefer:1998qe}.
However, entanglement is also a fragile diagnostic. Interactions with
unobserved fields, short-wavelength modes, reheating degrees of freedom, or
coarse-grained gravitational fluctuations can rapidly decohere the primordial
state \black{\cite{Kiefer:1998qe,Kiefer:2006je,Burgess:2014eoa,Schlosshauer:2014pgr,Zurek:2003zz}}. Once this
happens, the logarithmic negativity of a mode pair can vanish even though the
state still carries correlations in its covariance matrix. Thus the absence of
detectable entanglement does not necessarily mean that all traces of primordial
quantumness have disappeared.

A complementary covariance-based diagnostic is quantum discord. Discord captures nonclassical
correlations that can remain finite even in mixed separable states
\cite{Henderson:2001wrr,Modi:2012baj}. This distinction is especially natural
for cosmology, where the relevant perturbations are continuous-variable
Gaussian systems and their statistical information is encoded in covariance
matrices \cite{Weedbrook:2011wxo,Adesso:2014npz}. In this language, decoherence
does not simply switch correlations off. Instead, it reshapes the covariance
matrix: entanglement may disappear first, while anomalous phase-space coherence
and Gaussian discord can survive over a wider range of parameters. The question
is then not only whether primordial perturbations were quantum, but whether any
part of their residual covariance structure can be transferred to an observable
sector.

Scalar-induced gravitational waves provide a particularly clean setting in
which to ask this question. These gravitational waves are generated at second
order when scalar perturbations source tensor modes through the nonlinear
Einstein equations \cite{Ananda:2006af,Baumann:2007zm,Saito:2008jc,Domenech:2021ztg}.
They are widely studied in connection with enhanced small-scale scalar power,
primordial black-hole formation, and upcoming gravitational-wave searches
\cite{Carr:2021bzv,LISA:2017pwj,NANOGrav:2023gor,EPTA:2023fyk,Xu:2023wog}.
The important point for the present work is structural: the tensor source is
quadratic in the scalar perturbation. As a result, the induced tensor
covariance is determined by scalar four-point functions. For a Gaussian scalar
state these four-point functions factorize into Wick contractions, and
different contractions carry different physical information.
\black{Related SIGW calculations explicitly track the distinction between
connected and disconnected source correlators when non-Gaussian statistics are
present \cite{Adshead:2021ng,Perna:2024ng}.}

\begin{revision}
The central issue is more subtle than assigning one scalar contraction to the
power spectrum and another to a new observable.  For a real scalar field, all
normal and anomalous oscillator correlators enter the same basis-invariant
unequal-time Wightman function.  The induced tensor two-point function is
obtained from two Wick cross-contractions of this Wightman function.  We
therefore keep the scalar occupation $N_k^\zeta$ and anomalous correlator
$M_k^\zeta$ only as a bookkeeping device in a specified oscillator basis and
propagate the complete Wightman function through the second-order source.
This avoids introducing an additional, independent equal-time ``coherence
power spectrum'' for the real field.

The tensor field generated by a quadratic scalar source is non-Gaussian even
when the scalar state is Gaussian.  Its second moments can be summarized by an
out-mode occupation $d_k$ and anomalous moment $\gamma_k$, which define an
effective Gaussian reference state and a corresponding Gaussian discord.
However, the physical tensor four-point function also contains a connected
non-Gaussian cumulant that is not fixed by $d_k$ and $\gamma_k$.  Consequently,
the connected covariance of opposite-mode occupation is not generally equal
to $|\gamma_k|^2$.

The observable proposed here is therefore a baseline difference,
\begin{equation}
 \Delta\kappa_{\rm coh}(k)
 =\kappa_N[G^>_{\rm coh}](k)-\kappa_N[G^>_{\rm ref}](k),
 \label{eq:intro_coherence_excess}
\end{equation}
where both unequal-time Wightman functions and the corresponding scalar
parameters are fully specified.  A simple phase twirl in a fixed oscillator
basis sets $M_k^\zeta$ to zero while preserving $N_k^\zeta$, but it does not
in general preserve the equal-time field power because the $M_k^\zeta$ terms
also enter $G_k^>(\tau,\tau)$.  For an observational residual, the reference
model must therefore be recalibrated or marginalized so that it fits the same
diagonal scalar and SIGW two-point information.  The subtraction then removes
the connected four-point contribution already present for the ordinary
quadratic SIGW source and isolates dependence on the chosen phase-sensitive
unequal-time template.  It is not by itself an operational proof of
quantumness: a classical nonstationary source with the same measured
correlators can mimic it.  A genuine quantum certification would
require an additional nonclassicality criterion involving incompatible
observables, which is outside the scope of gravitational-wave strain
measurements considered here.

Figure~\ref{fig:scheme} summarizes this logically separated construction.  A
specified scalar Gaussian state and decoherence channel determine the unequal-
time Wightman function.  The retarded/in--in organization of the transfer follows
the standard Schwinger--Keldysh treatment of cosmological correlation functions
\cite{Weinberg:2005vy,Adshead:2009cb}.  The quadratic source then determines both tensor
second moments and a connected tensor cumulant.  The ordinary spectrum probes
the two-point sector, while a detector four-point statistic probes the
baseline-subtracted excess.  Any phase-sensitive second-moment observable also
requires an explicit phase reference and is strongly suppressed by ordinary
time averaging if no coherent demodulation is available.

The production of squeezed primordial perturbations is not new; it goes back
to Grishchuk and subsequent work on the quantum-to-classical transition
\cite{Grishchuk:1990bj,Albrecht:1992kf,Kiefer:1998jk,Martin:2015qta}.  The
novelty claimed here is deliberately narrower.  We (a) formulate the scalar
state in a fixed reference basis while using only the basis-invariant Wightman
function in the transfer calculation; (b) distinguish the effective Gaussian
tensor covariance from the exact non-Gaussian tensor state; (c) define a
baseline-subtracted four-point template; and (d) state explicitly which
observational numbers are phenomenological sensitivity proxies rather than
model predictions.  Recent LVK searches constrain the ordinary SIGW spectrum,
while dedicated work has begun to construct optimal gravitational-wave
four-point estimators \cite{LVK:2025O4a,Ciprini:2026pvz}.

Our main results are as follows.

\textbf{(i) Basis-consistent scalar-to-tensor transfer.}
The complete unequal-time scalar Wightman function, rather than separate
basis-dependent field spectra, is inserted in the retarded SIGW source.  This
provides the correct Gaussian-scalar tensor two-point function.

\textbf{(ii) Gaussian reference state versus exact tensor state.}
The moments $d_k$ and $\gamma_k$ define a physical Gaussian reference
covariance and its Gaussian discord, but the exact quadratic-source tensor
state is non-Gaussian and has an additional connected fourth cumulant.

\textbf{(iii) Baseline-subtracted connected observable.}
We define $\Delta\kappa_{\rm coh}$ by subtracting a fully specified
phase-randomized reference model, with the diagonal two-point sector refitted
when necessary.  This is the quantity that an exact scalar eight-point
calculation must predict before a detector claim can be made.

\textbf{(iv) Quantitative but explicitly proxy-level observational study.}
We compute the standard radiation-era SIGW two-point spectrum, compare its
normalization with representative LVK O1--O4a limits, and \black{evaluate
idealized sample-count and foreground-dilution scalings together with a
phenomenological Fisher illustration for representative detector bands. These
curves are not collaboration sensitivity forecasts.} The adopted four-point
amplitudes and network efficiencies are inputs.  The exact scalar eight-point
convolution and \black{four-channel network response} remain necessary for a
physical forecast.
\end{revision}

\begin{revision}
The paper is organized as follows.  Section~\ref{sec:framework} reviews the
Gaussian covariance tools.  Section~\ref{sec:scalar} fixes the scalar basis and
decoherence parameterizations.  Section~\ref{sec:SIGW_theory} states the tensor
normalization conventions, and Sec.~\ref{sec:tensor_discord} formulates the
unequal-time transfer and the effective tensor covariance.  Section~\ref{sec:observables}
defines the connected and phase-referenced diagnostics.  Section~\ref{sec:forecast}
gives the idealized four-point sensitivity scaling, while
Sec.~\ref{sec:numerical_consistency} presents the radiation-era SIGW
calculation and detector-weighted template benchmark.  We conclude in
Sec.~\ref{sec:conclusions}.
\end{revision}

\section{Theoretical Framework}
\label{sec:framework}

The problem studied in this paper naturally involves two ingredients. The first
is the phase-space description of primordial perturbations as continuous-variable
quantum systems. The second is the nonlinear scalar-to-tensor coupling that
allows information stored in scalar correlations to enter the induced
gravitational-wave sector. In this section we set up the first ingredient: the
Gaussian covariance-matrix language used to quantify entanglement, discord, and
opposite-mode coherence.

\begin{revision}
\begin{figure}[!htbp]
 \centering
 \includegraphics[width=0.90\textwidth]{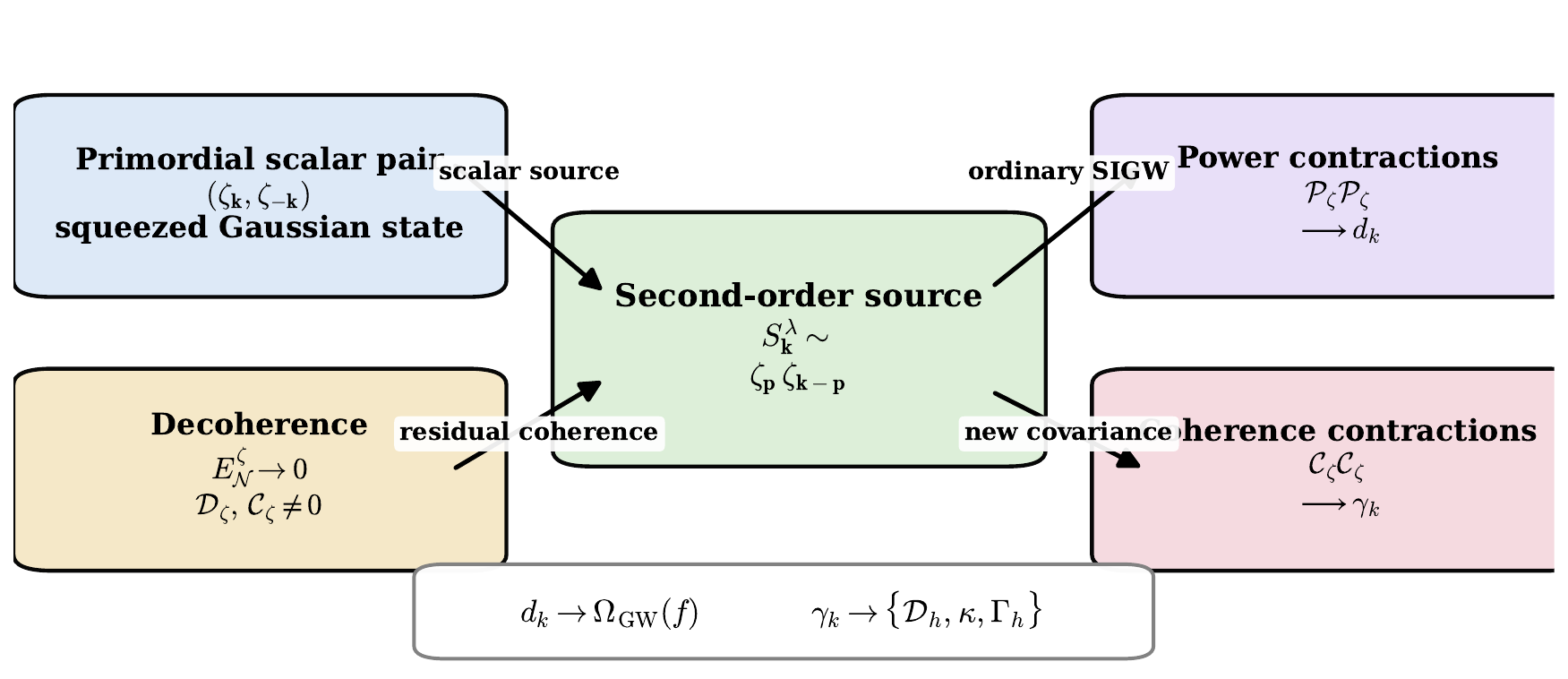}
 \caption{\black{Logical structure of the calculation.}  The basis-dependent
 oscillator moments $N_k^\zeta$ and $M_k^\zeta$ enter observables only through
 the complete unequal-time Wightman function.  The quadratic SIGW source
 produces tensor second moments and, independently, a connected non-Gaussian
 cumulant.  The ordinary spectrum is a two-point observable, whereas the
 residual-coherence target is a baseline-subtracted four-point statistic.
 Phase-sensitive second moments require a specified phase reference.}
 \label{fig:scheme}
\end{figure}
\end{revision}

\subsection{Gaussian states and quantum correlation measures}
\label{sec:gaussian_formalism}

For a set of $N$ bosonic modes we collect the canonical quadratures into the
phase-space vector
\begin{equation}
	\hat{\boldsymbol{\xi}}
	=
	\left(
	\hat q_1,\hat p_1,\ldots,\hat q_N,\hat p_N
	\right)^T .
	\label{eq:gauss_quad_vector}
\end{equation}
The quadratures obey
\begin{equation}
	[\hat{\xi}_i,\hat{\xi}_j]
	=
	i\,\Omega_{ij},
	\label{eq:gauss_CCR}
\end{equation}
where
\begin{equation}
	\Omega
	=
	\bigoplus_{n=1}^{N}
	\begin{pmatrix}
		0 & 1 \\
		-1 & 0
	\end{pmatrix}
	\label{eq:gauss_symplectic_form}
\end{equation}
is the symplectic form. This is the standard phase-space description of continuous-variable Gaussian quantum systems \cite{Weedbrook:2011wxo,Adesso:2014npz}. Throughout this work we use a discrete-mode normalization, appropriate for a finite comoving volume. The continuum limit is obtained by restoring the usual Dirac delta functions. We take the first moments to vanish,
\begin{equation}
	\langle \hat{\boldsymbol{\xi}}\rangle=0,
\end{equation}
since local displacements do not affect the correlation measures considered
below. The state is then characterized by the covariance matrix
\begin{equation}
	\sigma_{ij}
	=
	\frac12
	\left\langle
	\Delta\hat{\xi}_i\Delta\hat{\xi}_j
	+
	\Delta\hat{\xi}_j\Delta\hat{\xi}_i
	\right\rangle ,
	\qquad
	\Delta\hat{\xi}_i
	=
	\hat{\xi}_i-\langle\hat{\xi}_i\rangle .
	\label{eq:gauss_covariance_matrix}
\end{equation}
A covariance matrix represents a physical quantum state only if it satisfies
the Robertson--Schr\"odinger uncertainty condition
\begin{equation}
	\sigma+\frac{i}{2}\Omega\geq0 .
	\label{eq:gauss_RS_condition}
\end{equation}
Equivalently, all symplectic eigenvalues of $\sigma$ must obey
$\nu_\ell\geq1/2$. For a bipartite two-mode Gaussian state, the covariance matrix may be written
in block form,
\begin{equation}
	\sigma_{AB}
	=
	\begin{pmatrix}
		A & C\\
		C^T & B
	\end{pmatrix}.
	\label{eq:gauss_block_CM}
\end{equation}
The blocks $A$ and $B$ describe the local fluctuations of the two subsystems,
whereas $C$ contains their correlations \cite{Weedbrook:2011wxo,Adesso:2014npz}. In the applications below the two
subsystems will usually be the opposite-momentum modes
$(\mathbf{k},-\mathbf{k})$.

\begin{figure}[!htbp]
	\centering
	\includegraphics[width=0.85\textwidth]{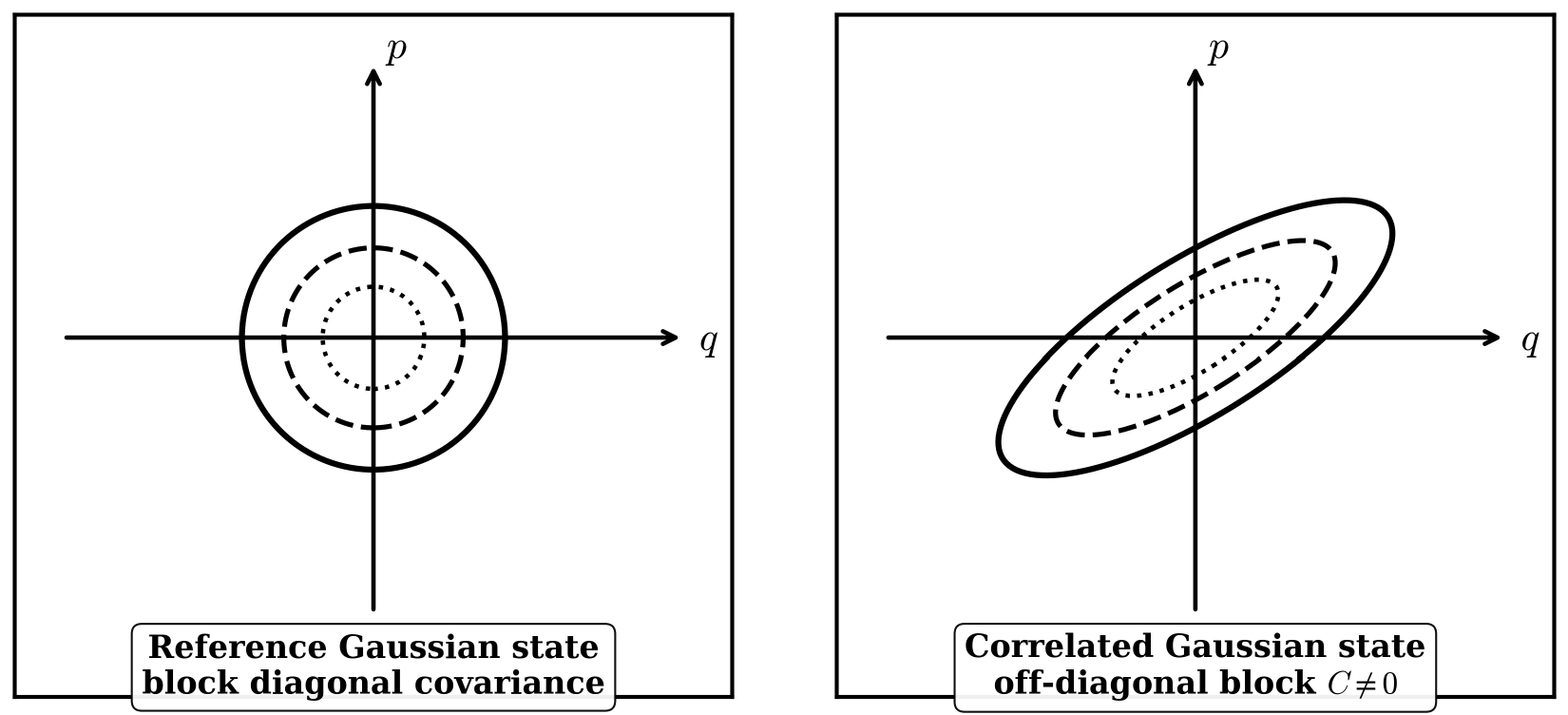}
	\caption{
		Phase-space representation of Gaussian covariance structure for a
		two-mode system $(\mathbf{k},-\mathbf{k})$. Left: an uncorrelated
		Gaussian reference state with a block-diagonal covariance matrix and
		vanishing inter-mode correlations. Right: a correlated Gaussian state
		in which the off-diagonal covariance block correlates the two modes.
		The deformation and orientation of the Wigner-function contours encode
		the strength and phase of the correlations, while the area of each
		contour is related to the mixedness of the state.
	}
	\label{fig:cov_geometry}
\end{figure}

A useful way to visualize this structure is shown in Fig.~\ref{fig:cov_geometry}. A block-diagonal Gaussian covariance matrix describes a reference state with no opposite-mode coherence, while a nonzero off-diagonal block encodes correlations between the two modes. In the following sections this same covariance-matrix language is used first for scalar perturbations and then for the induced tensor mode pair.

The symplectic spectrum is obtained from the invariants
\begin{equation}
	I_1=\det A,
	\qquad
	I_2=\det B,
	\qquad
	I_3=\det C,
	\qquad
	I_4=\det\sigma_{AB}.
	\label{eq:gauss_symplectic_invariants}
\end{equation}
For a two-mode Gaussian state,
\begin{equation}
	\nu_{\pm}
	=
	\sqrt{
		\frac{
			\Delta
			\pm
			\sqrt{\Delta^2-4I_4}
		}{2}
	},
	\qquad
	\Delta
	=
	I_1+I_2+2I_3 .
	\label{eq:gauss_symplectic_eigenvalues}
\end{equation}
The von Neumann entropy is then
\begin{equation}
	S(\sigma)
	=
	\sum_{\ell} f(\nu_\ell),
	\label{eq:gauss_entropy}
\end{equation}
with
\begin{equation}
	f(x)
	=
	\left(x+\frac12\right)\ln\left(x+\frac12\right)
	-
	\left(x-\frac12\right)\ln\left(x-\frac12\right).
	\label{eq:gauss_entropy_function}
\end{equation}
We use natural logarithms throughout.
The total correlation between the two subsystems is measured by the mutual
information,
\begin{equation}
	I(A:B)
	=
	S(A)+S(B)-S(AB).
	\label{eq:gauss_mutual_information}
\end{equation}
To separate the classical part, one considers Gaussian measurements on
subsystem $B$. The corresponding classical correlation is
\begin{equation}
	J(A|B)
	=
	S(A)
	-
	\inf_{\Pi_B^G}
	S(A|\Pi_B^G),
	\label{eq:gauss_classical_correlations}
\end{equation}
where the minimization is over Gaussian measurements. The Gaussian quantum
discord is then
\begin{equation}
	\mathcal{D}_{A|B}
	=
	I(A:B)-J(A|B).
	\label{eq:gauss_discord}
\end{equation}
This quantity captures nonclassical correlations that are not necessarily
entanglement \black{\cite{Henderson:2001wrr,Modi:2012baj,Giorda:2010wpy,Adesso:2010bfw}}. This distinction is important in cosmology, where environmental
decoherence can remove entanglement while leaving a mixed state with nonzero
discord. For comparison, we also use the logarithmic negativity. Partial transposition
with respect to subsystem $B$ changes the sign of one momentum quadrature in
that subsystem. At the level of the symplectic invariants this amounts to
\begin{equation}
	\Delta
	\longrightarrow
	\widetilde{\Delta}
	=
	I_1+I_2-2I_3 .
	\label{eq:gauss_PT_delta}
\end{equation}
The symplectic eigenvalues of the partially transposed covariance matrix are
\begin{equation}
	\widetilde{\nu}_{\pm}
	=
	\sqrt{
		\frac{
			\widetilde{\Delta}
			\pm
			\sqrt{\widetilde{\Delta}^2-4I_4}
		}{2}
	}.
	\label{eq:gauss_PT_eigenvalues}
\end{equation}
The logarithmic negativity is
\begin{equation}
	E_{\mathcal N}
	=
	\max
	\left[
	0,
	-\ln\left(2\widetilde{\nu}_{-}\right)
	\right].
	\label{eq:gauss_log_negativity}
\end{equation}
This form follows from the partial-transpose separability criterion for continuous-variable Gaussian states and from the logarithmic-negativity entanglement measure \black{\cite{Simon:2000zz,Duan:2000awi,Vidal:2002zz}}. A two-mode Gaussian state is entangled when $\widetilde{\nu}_{-}<1/2$. When $\widetilde{\nu}_{-}\geq1/2$, the logarithmic negativity vanishes and the state is separable. Separability, however, does not by itself imply the absence of discord. In the tensor sector, the covariance matrix should be built from canonical quadratures rather than directly from the metric amplitude. For each tensor polarization one may introduce a canonical variable of the form

\begin{equation}
	\hat v_{\mathbf{k}}^{\lambda}
	\propto
	aM_{\rm Pl}\hat h_{\mathbf{k}}^{\lambda},
	\qquad
	\hat\pi_{\mathbf{k}}^{\lambda}
	=
	\left(\hat v_{\mathbf{k}}^{\lambda}\right)' ,
	\label{eq:gauss_tensor_canonical_variables}
\end{equation}
with the proportionality factor fixed by the convention chosen for
$h_{ij}$ \cite{Mukhanov:1990me}. The Gaussian correlation measures introduced above are therefore
computed from the covariance matrix of
\begin{equation}
	\left(
	\hat v_{\mathbf{k}}^{\lambda},
	\hat\pi_{\mathbf{k}}^{\lambda},
	\hat v_{-\mathbf{k}}^{\lambda},
	\hat\pi_{-\mathbf{k}}^{\lambda}
	\right).
\end{equation}
Only after this canonical normalization is specified should the result be
translated into correlators of the gravitational-wave amplitude
$h_{\mathbf{k}}^{\lambda}$. The role of this section is thus limited but essential: it fixes the
phase-space notation and the correlation measures. The next section applies
these tools to primordial scalar perturbations, modeled as squeezed Gaussian
states subject to decoherence.

\section{Primordial curvature perturbations as Gaussian mode pairs}
\label{sec:scalar}
\begin{revision}
The curvature perturbation is expanded in a chosen normalized reference basis as
\begin{equation}
 \hat\zeta_{\mathbf k}(\tau)=u_k(\tau)\hat a_{\mathbf k}
 +u_k^*(\tau)\hat a_{-\mathbf k}^\dagger,
 \qquad
 [\hat a_{\mathbf k},\hat a_{\mathbf k'}^\dagger]
 =(2\pi)^3\delta^{(3)}(\mathbf k-\mathbf k').
 \label{eq:scalar_reference_expansion}
\end{equation}
For the Bunch--Davies vacuum in quasi-de Sitter space one may choose
\begin{equation}
 u_k^{\rm BD}(\tau)=
 \frac{H}{2M_{\rm Pl}\sqrt{\epsilon_{\rm sr}}\,k^{3/2}}
 (1+ik\tau)e^{-ik\tau},
 \label{eq:scalar_modefunc}
\end{equation}
which gives
\begin{equation}
 \mathcal P_\zeta^{\rm BD}(k)
 =\frac{k^3}{2\pi^2}|u_k^{\rm BD}|^2
 \xrightarrow{|k\tau|\ll1}
 \frac{H^2}{8\pi^2M_{\rm Pl}^2\epsilon_{\rm sr}}.
 \label{eq:scalar_power_spectrum}
\end{equation}

A basis convention is essential. Inflationary squeezing can be represented either by time-dependent Bogoliubov mode functions acting on a vacuum or by a squeezed state in a fixed reference basis. These are equivalent descriptions when the modes and operators are transformed together, but they must not be multiplied as two independent sources of squeezing. We use Eq.~\eqref{eq:scalar_reference_expansion} as the definition of the reference basis and encode the Gaussian state by its oscillator correlators. In the Bunch--Davies basis of Eq.~\eqref{eq:scalar_modefunc}, the Bunch--Davies vacuum itself has $N_k^\zeta=M_k^\zeta=0$. A nonzero pair $(N_k^\zeta,M_k^\zeta)$ in that same fixed basis therefore specifies a different Gaussian initial state; alternatively, the same physical Bunch--Davies state may acquire nonzero basis-dependent moments after a Bogoliubov change of oscillator basis. Under such a redefinition, $N_k^\zeta$ and $M_k^\zeta$ change, whereas the complete Wightman function below does not. Consequently, only a change in the full Wightman function relative to a stated physical baseline can generate a new prediction. \black{Equation~\eqref{eq:scalar_modefunc} fixes the inflationary reference basis; it is not, by itself, the radiation-era scalar transfer function. When the Wightman function is propagated to the SIGW production epoch, the post-inflationary scalar evolution must be included exactly once, either in the evolved mode functions $u_k(\tau)$ or in the source kernel defined below.} \black{Other non-vacuum choices can likewise generate additional correlation
structures. For example, coherent scalar initial states with a nonzero mean can
induce statistical inhomogeneity or anisotropy in the gravitational-wave
background \cite{Mukherjee:2025dcv}.}

For a pure two-mode squeezed state we adopt the sign convention
\begin{equation}
 |\Psi_{\rm sq}\rangle=\hat S(\xi_k)|0_{\mathbf k},0_{-\mathbf k}\rangle,
 \qquad
 \hat S(\xi_k)=
 \exp\!\left[
 \xi_k\hat a_{\mathbf k}^\dagger\hat a_{-\mathbf k}^\dagger
 -\xi_k^*\hat a_{\mathbf k}\hat a_{-\mathbf k}
 \right],
 \quad \xi_k=r_ke^{i\theta_k}.
 \label{eq:scalar_squeezing_operator}
\end{equation}
With this convention,
\begin{equation}
 N_k^\zeta\equiv\langle\hat a_{\mathbf k}^\dagger\hat a_{\mathbf k}\rangle
 =\sinh^2r_k,
 \qquad
 M_k^\zeta\equiv\langle\hat a_{\mathbf k}\hat a_{-\mathbf k}\rangle
 =e^{i\theta_k}\sinh r_k\cosh r_k.
 \label{eq:scalar_NM_pure}
\end{equation}
For a general homogeneous, isotropic, zero-mean Gaussian state,
\begin{align}
 \langle\hat a_{\mathbf k}^\dagger\hat a_{\mathbf k'}\rangle
 &=(2\pi)^3\delta^{(3)}(\mathbf k-\mathbf k')N_k^\zeta,\\
 \langle\hat a_{\mathbf k}\hat a_{\mathbf k'}\rangle
 &=(2\pi)^3\delta^{(3)}(\mathbf k+\mathbf k')M_k^\zeta,
 \label{eq:scalar_NM_continuum}
\end{align}
with the physicality condition
\begin{equation}
 |M_k^\zeta|^2\le N_k^\zeta(N_k^\zeta+1).
 \label{eq:scalar_NM_bound}
\end{equation}
The unequal-time Wightman function is
\begin{align}
 G_k^>(\tau,\tau')
 &\equiv\langle\hat\zeta_{\mathbf k}(\tau)
 \hat\zeta_{-\mathbf k}(\tau')\rangle \nonumber\\
 &=u_k(\tau)u_k^*(\tau')(1+N_k^\zeta)
 +u_k^*(\tau)u_k(\tau')N_k^\zeta \nonumber\\
 &\quad+u_k(\tau)u_k(\tau')M_k^\zeta
 +u_k^*(\tau)u_k^*(\tau')M_k^{\zeta *}.
 \label{eq:scalar_wightman_NM}
\end{align}
This is the basis-consistent object entering the SIGW source. At equal times, the terms involving $M_k^\zeta$ contribute to the same real-field power; they are not a second independent scalar power spectrum.
\begin{bluechange}
Explicitly, the equal-time dimensionless scalar spectrum associated with this
state is
\begin{equation}
 \mathcal P_\zeta(k,\tau)=\frac{k^3}{2\pi^2}
 \left[(2N_k^\zeta+1)|u_k(\tau)|^2
 +2\,\mathrm{Re}\!\left(M_k^\zeta u_k^2(\tau)\right)\right].
 \label{eq:scalar_equal_time_power}
\end{equation}
Thus the log-normal spectrum used later should be interpreted as the calibrated
equal-time (diagonal) input at the chosen primordial reference epoch. Any
coherent state specified by $(N_k^\zeta,M_k^\zeta)$ and any comparison state
must satisfy Eq.~\eqref{eq:scalar_NM_bound} and reproduce the intended diagonal
spectrum before their higher-order predictions are compared.
\end{bluechange}

For comparison with the covariance-matrix diagnostics, define normalized quadratures $\hat q=(\hat a+\hat a^\dagger)/\sqrt2$ and $\hat p=(\hat a-\hat a^\dagger)/(i\sqrt2)$. After a local phase rotation that makes $M_k^\zeta$ real and non-negative, a symmetric two-mode covariance matrix can be written
\begin{equation}
 \sigma_\zeta=
 \begin{pmatrix}
 a_k&0&c_k&0\\
 0&a_k&0&-c_k\\
 c_k&0&a_k&0\\
 0&-c_k&0&a_k
 \end{pmatrix},
 \qquad
 a_k=N_k^\zeta+\frac12,
 \quad c_k=|M_k^\zeta|.
 \label{eq:scalar_decohered_CM}
\end{equation}
For the illustrative additive-noise channel used in the covariance plots,
\begin{equation}
 a_k=\frac12\cosh(2r_k)+\epsilon_{\rm dec},
 \qquad
 c_k=\frac12\sinh(2r_k),
 \qquad \epsilon_{\rm dec}\ge0.
 \label{eq:scalar_a_c_definitions}
\end{equation}
Its symplectic eigenvalues and smallest partially-transposed eigenvalue are
\begin{equation}
 \nu_k=\sqrt{a_k^2-c_k^2}
 =\sqrt{\frac14+\epsilon_{\rm dec}\cosh(2r_k)+\epsilon_{\rm dec}^2},
 \qquad
 \widetilde\nu_-=a_k-c_k=\frac12e^{-2r_k}+\epsilon_{\rm dec}.
 \label{eq:scalar_symplectic_PT}
\end{equation}
Thus
\begin{equation}
 E_{\mathcal N}^\zeta=\max\{0,-\ln(2\widetilde\nu_-)\},
 \qquad
 \epsilon_{\rm dec}<\frac{1-e^{-2r_k}}{2}
 \quad\hbox{for entanglement}.
 \label{eq:scalar_logneg_threshold}
\end{equation}
For this symmetric squeezed-thermal form, heterodyne detection is the
optimal Gaussian measurement and the Gaussian discord is
\black{\cite{Giorda:2010wpy,Adesso:2010bfw}}
\begin{equation}
 \mathcal D_\zeta
 =f(a_k)-2f(\nu_k)
 +f\!\left(a_k-\frac{c_k^2}{a_k+1/2}\right),
 \label{eq:scalar_discord_symmetric}
\end{equation}
where
\begin{equation}
 f(x)=\left(x+\frac12\right)\ln\left(x+\frac12\right)
 -\left(x-\frac12\right)\ln\left(x-\frac12\right).
\end{equation}
The dimensionless anomalous-correlation fraction is
\begin{equation}
 \chi_k\equiv\frac{|M_k^\zeta|}{N_k^\zeta+1/2}=\frac{c_k}{a_k},
 \qquad 0\le\chi_k\le1.
 \label{eq:scalar_chi_dec}
\end{equation}
The additive channel in Eq.~\eqref{eq:scalar_a_c_definitions} increases the diagonal variance without directly damping $M_k^\zeta$. A rate equation that damps $M_k^\zeta$ is a different Gaussian channel and is introduced separately below.

\begin{figure}[!htbp]
 \centering
 \includegraphics[width=0.92\textwidth]{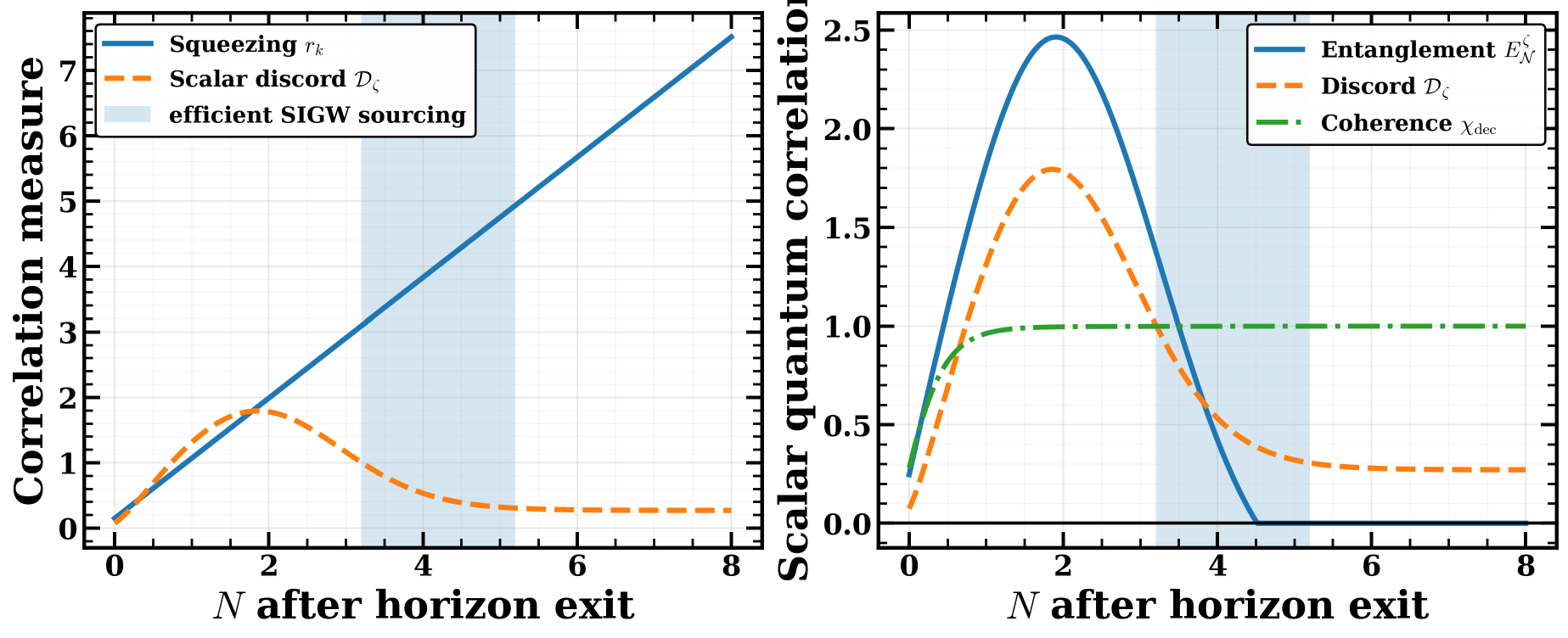}
 \caption{Covariance-level illustration of squeezing, logarithmic negativity, Gaussian discord, and $\chi_k$ in the additive-noise model. The figure is a diagnostic of that specified Gaussian channel, not a microscopic decoherence-rate calculation. \black{Here $r_k$ parameterizes the state covariance in the fixed oscillator basis; it must not be applied again as an independent Bogoliubov squeezing of the Bunch--Davies mode function.}}
 \label{fig:squeezing_discord}
\end{figure}
\end{revision}
\section{Scalar-induced gravitational waves and normalization conventions}
\label{sec:SIGW_theory}
\begin{revision}
\black{We work in Poisson (Newtonian) gauge and use}
\begin{equation}
 ds^2=a^2(\tau)\left[-(1+2\Phi)d\tau^2+
 \big((1-2\Psi)\delta_{ij}+h_{ij}\big)dx^idx^j\right],
 \qquad \partial_i h_{ij}=h_{ii}=0,
 \label{eq:sigw_metric}
\end{equation}
with no additional factor $1/2$ multiplying $h_{ij}$. The polarization tensors obey $e_{ij}^\lambda e_{ij}^{\lambda'}=\delta^{\lambda\lambda'}$, and
\begin{equation}
 h_{ij}(\tau,\mathbf x)=\sum_\lambda\int\frac{d^3k}{(2\pi)^3}
 e^{i\mathbf k\cdot\mathbf x}e_{ij}^\lambda(\mathbf k)
 h_{\mathbf k}^\lambda(\tau).
 \label{eq:sigw_polarization_decomposition}
\end{equation}
The second-order equation is written in convention-independent kernel notation as
\begin{equation}
 h_{\mathbf k}^{\lambda\prime\prime}+2\mathcal Hh_{\mathbf k}^{\lambda\prime}
 +k^2h_{\mathbf k}^\lambda=S_{\mathbf k}^\lambda,
 \qquad
 S_{\mathbf k}^\lambda(\tau)=
 \int\frac{d^3p}{(2\pi)^3}\,
 \mathcal Q_\lambda(\mathbf k,\mathbf p;\tau)
 \zeta_{\mathbf p}(\tau)\zeta_{\mathbf k-\mathbf p}(\tau).
 \label{eq:sigw_source_generic}
\end{equation}
\black{In the operator form of Eq.~\eqref{eq:sigw_source_generic}, $\mathcal Q_\lambda$ contains the polarization projection, background/derivative coefficients, and normalization factors, while any scalar transfer already included in $\hat\zeta_{\mathbf p}(\tau)$ must not be inserted a second time. Equivalently, in the standard numerical formulation one may factor the scalar evolution out of the field operators and absorb it into the radiation-era $u$--$v$ kernel. We adopt the latter convention for the ordinary spectrum in Sec.~\ref{sec:numerical_consistency}; equivalent semi-analytic formulations are given in Refs.~\cite{Kohri:2018awv,Inomata:2018epa,Domenech:2021ztg}. For that standard kernel we assume negligible scalar anisotropic stress, so that $\Phi=\Psi$, and radiation domination when the modes of interest re-enter the Hubble radius. A non-standard post-inflationary equation of state or appreciable anisotropic stress changes the scalar transfer functions and therefore the induced spectrum \cite{Domenech:2021ztg}.} In a fundamental operator treatment, $\hat S_{\mathbf k}^\lambda$ is a composite operator and requires a stated renormalization prescription. The numerical calculation performed here instead starts from the finite, coarse-grained primordial spectrum $\mathcal P_\zeta$ used in the standard stochastic SIGW calculation; zero-point contributions are not included in that phenomenological input.

The retarded induced solution is
\begin{equation}
 \hat h_{\mathbf k,{\rm ind}}^\lambda(\tau)=
 \int_{\tau_i}^{\tau}d\tau'\,G_k^R(\tau,\tau')
 \hat S_{\mathbf k}^\lambda(\tau'),
 \label{eq:sigw_h_solution}
\end{equation}
where
\begin{equation}
 \left[\partial_\tau^2+2\mathcal H\partial_\tau+k^2\right]
 G_k^R(\tau,\tau')=\delta(\tau-\tau'),
 \qquad G_k^R(\tau,\tau')=0\quad(\tau<\tau').
 \label{eq:sigw_green_eq}
\end{equation}
The full canonical tensor field, not the sourced particular solution alone, carries the canonical commutator. With $e_{ij}^\lambda e_{ij}^{\lambda'}=\delta^{\lambda\lambda'}$, the quadratic action is \cite{Mukhanov:1990me,Domenech:2021ztg}
\begin{align}
 S_h^{(2)}&=\frac{M_{\rm Pl}^2}{8}\int d\tau\,d^3x\,a^2\left[(h_{ij}')^2-(\partial_\ell h_{ij})^2\right]\nonumber\\
 &=\frac12\sum_\lambda\int d\tau\,\frac{d^3k}{(2\pi)^3}\,\left[|v_{\mathbf k}^{\lambda\prime}|^2-\left(k^2-\frac{a''}{a}\right)|v_{\mathbf k}^\lambda|^2\right],
 \label{eq:sigw_tensor_action}
\end{align}
where, for each polarization,
\begin{equation}
 \hat v_{\mathbf k}^\lambda=\frac{aM_{\rm Pl}}{2}\hat h_{\mathbf k}^\lambda,
 \qquad \hat\pi_{\mathbf k}^\lambda=(\hat v_{\mathbf k}^\lambda)',
 \label{eq:sigw_canonical_tensor_variable}
\end{equation}
and, after the source has switched off and for modes in an adiabatic subhorizon regime, we choose a late-time positive-frequency basis and expand
\begin{equation}
 \hat v_{\mathbf k}^\lambda(\tau)=v_k^{\rm out}(\tau)\hat b_{\mathbf k}^\lambda
 +v_k^{{\rm out}*}(\tau)\hat b_{-\mathbf k}^{\lambda\dagger},
 \qquad
 v_k^{\rm out}v_k^{{\rm out}*\prime}
 -v_k^{{\rm out}\prime}v_k^{{\rm out}*}=i.
 \label{eq:tensor_out_expansion}
\end{equation}
The operators $\hat b_{\mathbf k}^\lambda$ are canonical operators in this chosen late-time adiabatic basis. In an expanding spacetime this particle interpretation is not globally unique; $d_k$ and $\gamma_k$ are therefore basis-dependent diagnostics, whereas detector strain correlators obtained after propagating the full field are the operational observables. Induced contributions are quoted with the homogeneous-vacuum contribution subtracted when necessary.
\black{At leading order we also assume a factorized primordial scalar--tensor state with no initial scalar--tensor cross-correlations. Under this assumption the homogeneous tensor vacuum does not generate interference terms with the induced scalar-sourced contribution in the induced second moments; relaxing it would require the corresponding cross-correlators to be retained.}

For each polarization,
\begin{equation}
 \langle h_{\mathbf k}^\lambda h_{\mathbf k'}^{\lambda'}\rangle
 =(2\pi)^3\delta^{(3)}(\mathbf k+\mathbf k')\delta^{\lambda\lambda'}
 \frac{2\pi^2}{k^3}\mathcal P_h^\lambda(k),
 \qquad
 \mathcal P_h=\sum_\lambda\mathcal P_h^\lambda.
 \label{eq:sigw_tensor_power}
\end{equation}
With these conventions, the oscillation-averaged subhorizon energy-density fraction is
\begin{equation}
 \Omega_{\rm GW}(\tau,k)=\frac1{12}
 \left(\frac{k}{\mathcal H}\right)^2
 \overline{\mathcal P_h(\tau,k)}.
 \label{eq:sigw_energy_density}
\end{equation}
The factor $1/12$ applies to the polarization-summed spectrum and unit-normalized polarization tensors defined above; the relation between strain power and gravitational-wave energy density is reviewed in Refs.~\cite{Domenech:2021ztg,Caprini:2018mtu,Romano:2016dpx}. Conventions with $e_{ij}^\lambda e_{ij}^{\lambda'}=2\delta^{\lambda\lambda'}$ or an explicit factor $1/2$ multiplying $h_{ij}$ redistribute factors between $h_\lambda$, $\mathcal P_h$, and the energy-density formula; these conventions must not be mixed.
\end{revision}
\section{Tensor out-mode covariance and unequal-time transfer}
\label{sec:tensor_discord}
\begin{revision}
Define the translationally invariant scalar two-point function by
\begin{equation}
 \langle\hat\zeta_{\mathbf p}(\tau_1)\hat\zeta_{\mathbf q}(\tau_2)\rangle
 =(2\pi)^3\delta^{(3)}(\mathbf p+\mathbf q)G_p^>(\tau_1,\tau_2),
 \label{eq:scalar_G_momentum}
\end{equation}
with $G_p^>$ given by Eq.~\eqref{eq:scalar_wightman_NM}. Substitution into the retarded particular solution gives
\begin{align}
 &\langle\hat h_{\mathbf k,{\rm ind}}^\lambda(\tau_o)
 \hat h_{\mathbf k',{\rm ind}}^{\lambda'}(\tau_o)\rangle_c \nonumber\\
 &=\int d\tau_1d\tau_2
 \int\frac{d^3p\,d^3p'}{(2\pi)^6}
 \mathcal I_\lambda(\mathbf k,\mathbf p;\tau_1)
 \mathcal I_{\lambda'}(\mathbf k',\mathbf p';\tau_2)\nonumber\\
 &\quad\times\Big[
 \langle\zeta_{\mathbf p}(\tau_1)\zeta_{\mathbf p'}(\tau_2)\rangle
 \langle\zeta_{\mathbf k-\mathbf p}(\tau_1)
 \zeta_{\mathbf k'-\mathbf p'}(\tau_2)\rangle\nonumber\\
 &\qquad\quad+
 \langle\zeta_{\mathbf p}(\tau_1)
 \zeta_{\mathbf k'-\mathbf p'}(\tau_2)\rangle
 \langle\zeta_{\mathbf k-\mathbf p}(\tau_1)
 \zeta_{\mathbf p'}(\tau_2)\rangle
 \Big],
 \label{eq:tensor_unequal_time_transfer}
\end{align}
where $\mathcal I_\lambda=G_k^R\mathcal Q_\lambda$ and the subscript $c$ removes the product of one-point source expectations. The momentum delta functions imply $\mathbf k'=-\mathbf k$. Equation~\eqref{eq:tensor_unequal_time_transfer} is the correct Gaussian-scalar starting point; no independent ``anomalous field power spectrum'' is introduced.

The out-mode occupation and anomalous correlator are
\begin{equation}
 d_k\equiv\langle\hat b_{\mathbf k}^{\lambda\dagger}\hat b_{\mathbf k}^\lambda\rangle,
 \qquad
 \gamma_k\equiv\langle\hat b_{\mathbf k}^\lambda\hat b_{-\mathbf k}^\lambda\rangle.
 \label{eq:tensor_out_correlators}
\end{equation}
They are obtained by first adding the homogeneous solution to the induced particular solution and then applying the Klein--Gordon projection associated with Eq.~\eqref{eq:tensor_out_expansion} to the full tensor correlators. Symbolically,
\begin{equation}
 d_k=\mathcal F_N[N^\zeta,M^\zeta;u,\mathcal I],
 \qquad
 \gamma_k=\mathcal F_M[N^\zeta,M^\zeta;u,\mathcal I].
 \label{eq:gammak_explicit_transfer}
\end{equation}
Vacuum, occupation, anomalous, and interference terms all occur when Eq.~\eqref{eq:scalar_wightman_NM} is expanded; a universal separation into $\mathcal P_\zeta\mathcal P_\zeta$ and an independent $\mathcal C_\zeta\mathcal C_\zeta$ is not valid.

The covariance matrix sharing the same second moments is
\begin{equation}
 \sigma_h^{\rm G}(k)=
 \begin{pmatrix}
 \alpha_k&0&\Re\gamma_k&\Im\gamma_k\\
 0&\alpha_k&\Im\gamma_k&-\Re\gamma_k\\
 \Re\gamma_k&\Im\gamma_k&\alpha_k&0\\
 \Im\gamma_k&-\Re\gamma_k&0&\alpha_k
 \end{pmatrix},
 \qquad \alpha_k=d_k+\frac12.
 \label{eq:tensor_CM_standard}
\end{equation}
It is physical when
\begin{equation}
 |\gamma_k|^2\le d_k(d_k+1),
 \label{eq:tensor_gamma_bound}
\end{equation}
which is the standard uncertainty-condition bound for a symmetric two-mode Gaussian covariance matrix \cite{Isar:2013vvd}, and its symplectic eigenvalue is $\nu_h=\sqrt{\alpha_k^2-|\gamma_k|^2}$. \black{This uncertainty bound applies to the full physical second moments. Vacuum-subtracted or ``induced-only'' pieces are bookkeeping contributions and need not, by themselves, define a positive covariance matrix; the bound should not be imposed on such subtracted pieces unless a genuine state with those moments has been specified.} The corresponding Gaussian-reference discord is
\begin{equation}
 \mathcal D_h^{\rm G}=f(\alpha_k)-2f(\nu_h)
 +f\!\left(\alpha_k-\frac{|\gamma_k|^2}{\alpha_k+1/2}\right).
 \label{eq:tensor_discord_correct}
\end{equation}
Because $h\sim\zeta^2$, the exact induced tensor state is generally non-Gaussian even when the scalar state is Gaussian. Equation~\eqref{eq:tensor_CM_standard} therefore defines the unique Gaussian state with the same second moments; it is not the complete tensor density operator. \black{This distinction between the covariance sector and connected higher-order SIGW statistics is also manifest in explicit non-Gaussian SIGW calculations \cite{Adshead:2021ng,Perna:2024ng}.}

For a phenomenological damping channel, let $N\equiv\ln(a/a_0)$ increase with cosmic time and define
\begin{equation}
 \frac{dM_k^\zeta}{dN}=-\frac{\Gamma_{\rm dec}(k,N)}{H(N)}M_k^\zeta,
 \qquad
 M_k^\zeta(N)=M_{k,0}^\zeta e^{-D_k(N)},
 \qquad
 D_k(N)=\int_{N_0}^{N}dN'\frac{\Gamma_{\rm dec}(k,N')}{H(N')}.
 \label{eq:integrated_decoherence}
\end{equation}
If, for a specified narrow feature and transfer kernel, the leading contribution to $\gamma_k$ contains exactly two factors of $M^\zeta$ while $d_k$ is unchanged to the required accuracy, then
\begin{equation}
 \gamma_k\simeq\gamma_{k,0}e^{-2D_k},
 \qquad
 \kappa_N^{\rm G}\equiv\frac{|\gamma_k|^2}{d_k^2}
 \simeq\kappa_{N,0}^{\rm G}e^{-4D_k}.
 \label{eq:decoherence_kappa_scaling}
\end{equation}
This is a conditional scaling law, not a universal identity. It additionally requires either that the $M_k^\zeta$-dependent correction to $d_k$ is negligible or that the diagonal two-point model is recalibrated while $D_k$ is varied. A microscopic master equation may also modify $N_k^\zeta$, the phase of $M_k^\zeta$, and the transfer kernel.

\begin{figure}[!htbp]
 \centering
 \includegraphics[width=0.92\textwidth]{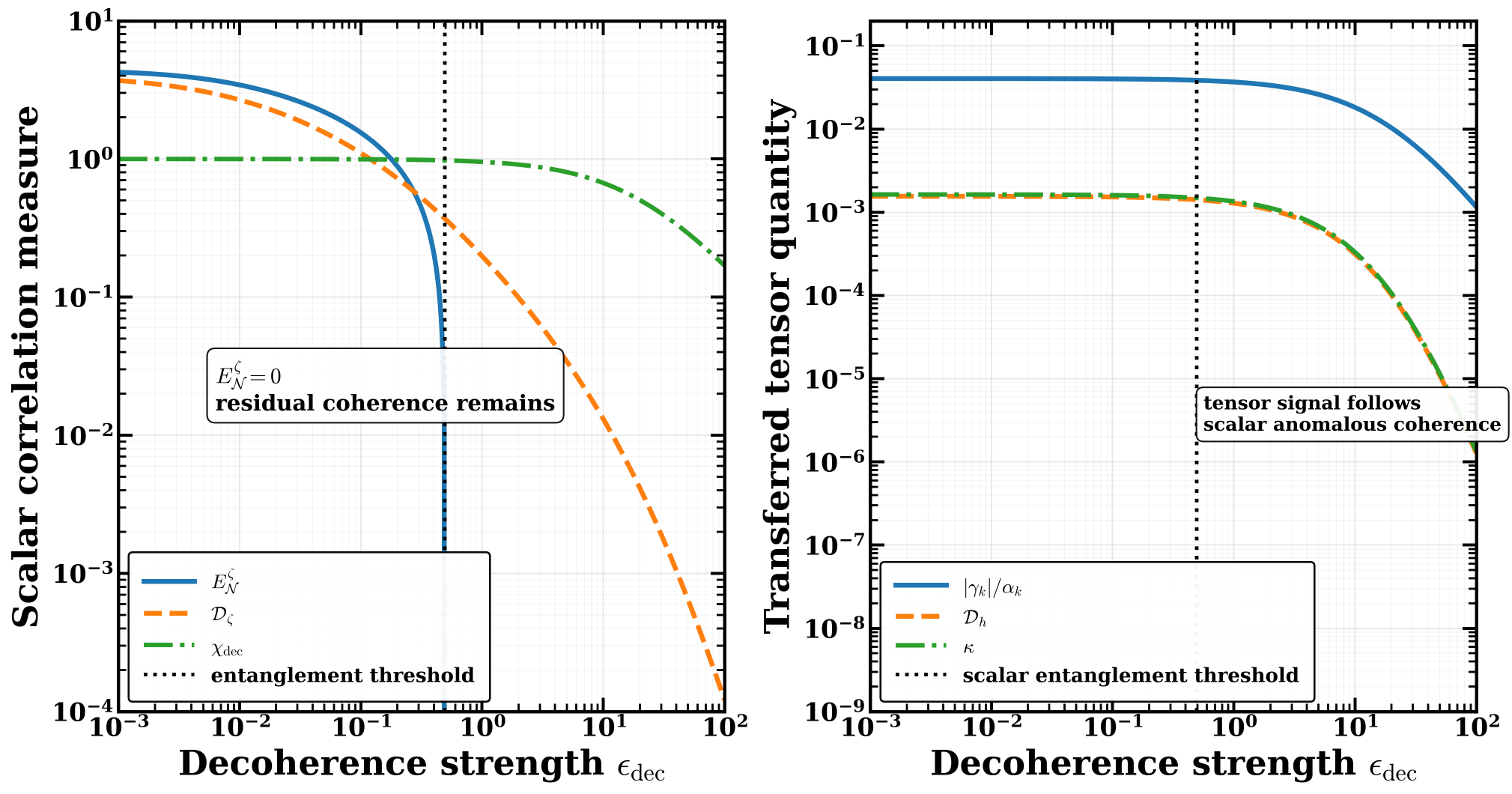}
 \caption{Illustrative scalar-to-tensor correlation transfer as a function of the additive-noise parameter $\epsilon_{\rm dec}$. The left panel displays the loss of scalar entanglement together with the persistence of Gaussian discord and the normalized anomalous correlation $\chi_k$. The right panel shows the corresponding Gaussian-reference tensor quantities. The curves are covariance-level benchmarks with a fixed transfer normalization; they are not a microscopic calculation of the rate $\Gamma_{\rm dec}$.}
 \label{fig:transfer_decoherence}
\end{figure}

The figure should therefore be read together with Eqs.~\eqref{eq:scalar_a_c_definitions} and \eqref{eq:integrated_decoherence}: $\epsilon_{\rm dec}$ characterizes the additive covariance channel used for visualization, whereas $D_k$ characterizes a distinct phase-damping channel. No universal mapping between these two parameters is assumed.

\end{revision}
\section{Observable signatures}
\label{sec:observables}
\begin{revision}
The ordinary SIGW spectrum is fixed by the tensor two-point function and Eq.~\eqref{eq:sigw_energy_density}. \black{Higher-order and non-Gaussian stochastic-background statistics have been developed as complementary probes beyond the ordinary two-point search, including fourth-order detector correlations, likelihood-based non-Gaussian searches, and connected-trispectrum estimators \cite{Seto:2009ng,Thrane:2013ng,Romano:2016dpx,Ciprini:2026pvz}.} Higher-order information in the present construction must be defined with operators that are independent for $\mathbf k$ and $-\mathbf k$. Let
\begin{equation}
 \hat N_{\mathbf k}=\hat b_{\mathbf k}^\dagger\hat b_{\mathbf k},
 \qquad
 \kappa_N(k)=
 \frac{\langle\hat N_{\mathbf k}\hat N_{-\mathbf k}\rangle
 -\langle\hat N_{\mathbf k}\rangle\langle\hat N_{-\mathbf k}\rangle}
 {d_k^2},
 \qquad d_k>0.
 \label{eq:obs_kappa_def}
\end{equation}
Second-order intensity/occupation correlations of this type are the gravitational-wave analogue of Hanbury Brown--Twiss observables \cite{Giovannini:2010xg,Toccacelo:2026hcz}. For a general zero-mean state,
\begin{equation}
 {\rm Cov}(\hat N_{\mathbf k},\hat N_{-\mathbf k})
 =|\gamma_k|^2+|\eta_k|^2+C_{N,k}^{(4)},
 \qquad
 \eta_k\equiv\langle\hat b_{\mathbf k}^\dagger\hat b_{-\mathbf k}\rangle,
 \label{eq:obs_general_number_covariance}
\end{equation}
where $C_{N,k}^{(4)}$ is the connected fourth-order cumulant beyond Gaussian Wick contractions. Statistical homogeneity gives $\eta_k=0$ for nonzero $\mathbf k$ in the continuum. Hence
\begin{equation}
 \kappa_N(k)=\underbrace{\frac{|\gamma_k|^2}{d_k^2}}_{\kappa_N^{\rm G}(k)}
 +\frac{C_{N,k}^{(4)}}{d_k^2}.
 \label{eq:obs_kappa_complete}
\end{equation}
The first term is the Gaussian-reference contribution. The second is generically nonzero for SIGWs because the tensor field is quadratic in the scalar field. The appropriate residual-coherence observable is consequently
\begin{equation}
 \Delta\kappa_{\rm coh}(k)
 \equiv\kappa_N[G^>_{\rm coh}](k)-\kappa_N[G^>_{\rm ref}](k),
 \label{eq:obs_coherence_excess}
\end{equation}
where the reference Wightman function is specified independently and, for a
fair observational comparison, its diagonal parameters are refitted to the
same scalar and tensor two-point data.  The shorthand $M_k^\zeta\to0$ at fixed
$N_k^\zeta$ is only a fixed-basis state-space diagnostic, because it generally
changes the equal-time scalar power.  Equation~\eqref{eq:obs_coherence_excess}
subtracts the ordinary quadratic-source baseline. A nonzero un-subtracted
trispectrum is not, by itself, evidence for primordial quantum coherence.

For the Gaussian reference state, Eq.~\eqref{eq:tensor_gamma_bound} gives
\begin{equation}
 0\le\kappa_N^{\rm G}\le1+\frac1{d_k}.
 \label{eq:kappa_gaussian_bound}
\end{equation}
Thus $\kappa_N^{\rm G}$ is not universally bounded by unity; the upper bound approaches one only in the high-occupation limit $d_k\gg1$.

The phase of the anomalous second moment may be probed only after a positive-frequency decomposition and a phase reference are specified. Defining
\begin{equation}
 \Gamma_h(k,\tau)=
 \langle\hat h_{\mathbf k}^{(+)}(\tau)\hat h_{-\mathbf k}^{(+)}(\tau)\rangle
 =\left(\frac{2v_k^{\rm out}(\tau)}{aM_{\rm Pl}}\right)^2\gamma_k,
 \label{eq:obs_Gamma_gamma}
\end{equation}
shows explicitly that the known propagation phase must be removed before $\arg\gamma_k$ can be inferred. For $v_k^{\rm out}\propto e^{-i\omega_k\tau}$, an un-demodulated average over a duration $T$ gives
\begin{equation}
 \frac1T\int_{-T/2}^{T/2}dt\,e^{-2i\omega_kt}
 =\frac{\sin(\omega_kT)}{\omega_kT}
 \equiv {\rm sinc}(2\pi fT),
 \label{eq:phase_averaging_sinc}
\end{equation}
so ordinary long-time averaging suppresses this quantity unless a coherent phase reference and demodulation prescription are supplied. A strictly stationary, phase-random reference background has $\gamma_k=0$; a nonzero value requires a time origin or external phase convention and can also be mimicked by classical nonstationary sources.

\begin{figure}[!htbp]
 \centering
 \includegraphics[width=0.88\textwidth]{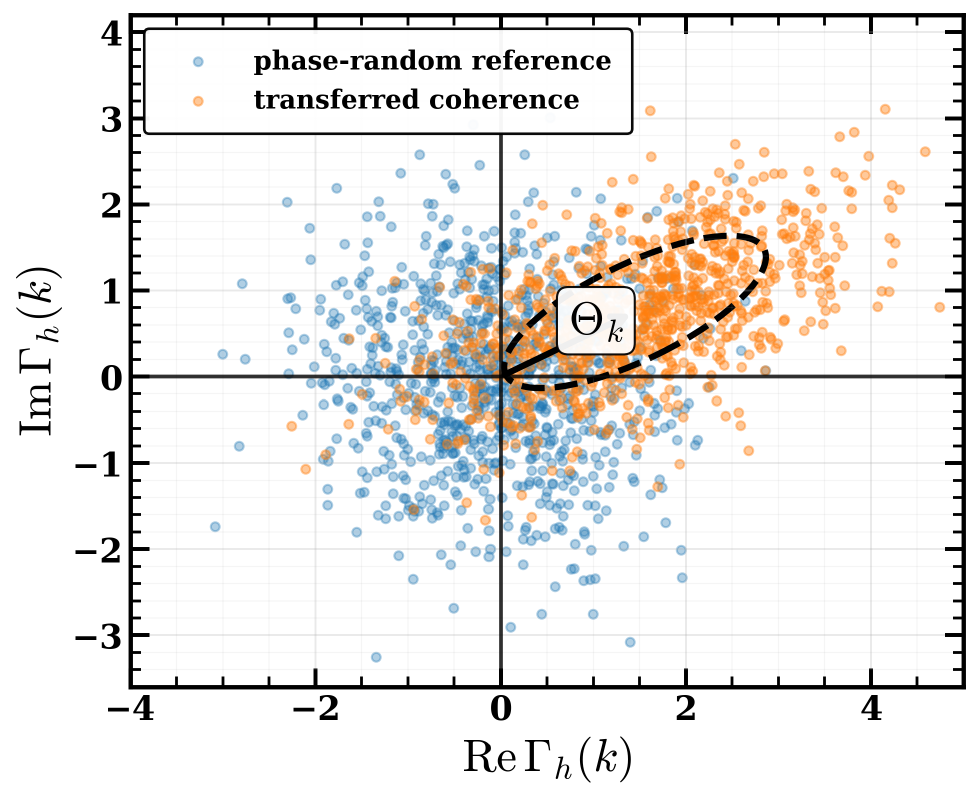}
 \caption{Illustrative complex-plane representation of a phase-sensitive correlator. The phase-random reference cloud is centered around vanishing coherent phase, whereas the transferred-coherence example is displaced and oriented along a preferred phase direction. This is a visualization of the information carried by $\arg\gamma_k$, not simulated detector data. In a real search, the propagation phase, detector response, and finite-time averaging in Eq.~\eqref{eq:phase_averaging_sinc} must be included explicitly.}
 \label{fig:phase_discrimination}
\end{figure}

\black{For the spectral illustrations we use the localized log-normal template adopted for the numerical benchmarks.} A log-normal scalar peak of this form was introduced and analytically studied for scalar-induced gravitational waves by Pi and Sasaki~\cite{PiSasaki:2020lognormal}, who derived useful analytic expressions for the induced spectrum in both the narrow- and broad-peak regimes,
\begin{equation}
 \mathcal P_\zeta(k)=A_\zeta
 \exp\!\left[-\frac{\ln^2(k/k_*)}{2\sigma_\zeta^2}\right],
 \label{eq:lognormal_peak_template}
\end{equation}
where $A_\zeta$ denotes the peak amplitude in this convention. This definition is distinct from the area-normalized LVK convention introduced later in Eq.~\eqref{eq:lvk_lognormal}.

\begin{figure}[!htbp]
 \centering
 \includegraphics[width=0.92\textwidth]{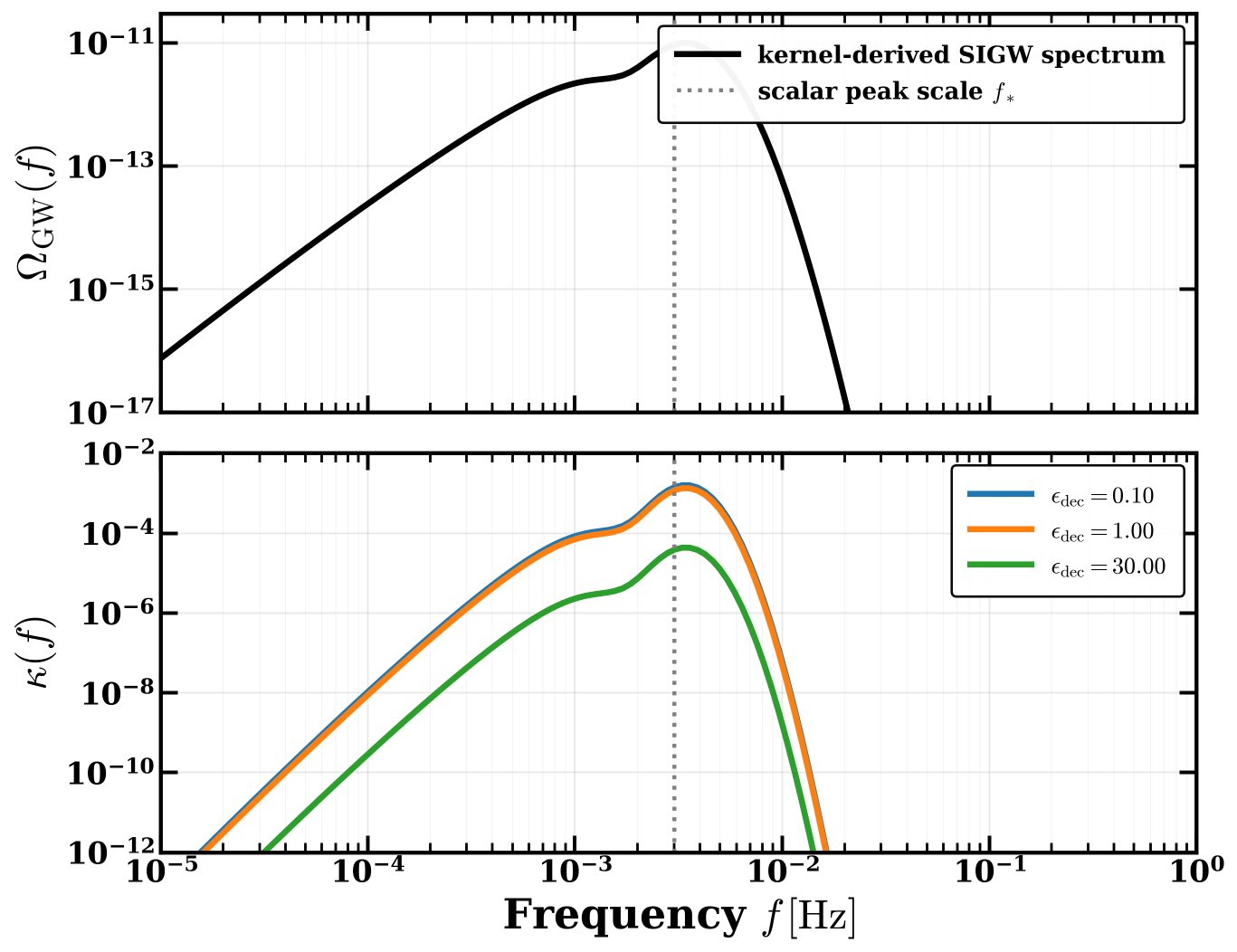}
 \caption{\black{Combined benchmark} for the localized scalar spectrum in Eq.~\eqref{eq:lognormal_peak_template}. Upper: the standard radiation-era SIGW spectrum. Lower: a coherence-dependent connected-covariance template for representative decoherence strengths. The lower panel illustrates the frequency localization and damping expected from the transferred-coherence sector; it should not be interpreted as the complete scalar eight-point SIGW trispectrum.}
 \label{fig:kernel_signals}
\end{figure}

A gravitational-wave interferometer measures strain rather than the mode occupation operator $\hat N_{\mathbf k}$. Moreover, for a real time series, negative temporal frequency is the complex conjugate of positive temporal frequency; it should not be identified with the independent spatial direction $-\hat{\mathbf n}$. A detector-level implementation of $\Delta\kappa_{\rm coh}$ must therefore project the angular tensor trispectrum into four strain channels with their antenna patterns and time delays. The occupation-space quantities above are theoretical diagnostics, while the measurable object is the corresponding network strain four-point function.

The physically meaningful joint target is therefore the consistency of the ordinary spectrum, the baseline-subtracted connected statistic $\Delta\kappa_{\rm coh}$, and any phase-referenced correlator under one specified scalar state, transfer kernel, and decoherence channel.
\end{revision}
\section{Observational prospects and four-point sensitivity}
\label{sec:forecast}
\begin{revision}
Future experiments probe complementary frequency bands: LISA and TianQin target the millihertz regime \black{\cite{LISA:2017pwj,TianQin:2015yph}}, DECIGO and BBO extend sensitivity toward the decihertz band \cite{Kawamura:2020pcg,Crowder:2005nr}, while pulsar-timing arrays probe nanohertz frequencies \cite{Burke-Spolaor:2018bvk,Romano:2016dpx}. These bands correspond to different scalar peak scales and therefore test complementary portions of the SIGW parameter space. The higher-order problem is more demanding than the ordinary stochastic-background search because unresolved astrophysical populations and instrumental correlations contribute both to the variance and, in general, to connected cumulants.

\subsection{Normalization to current SIGW limits}
The LVK O1--O4a analysis used an area-normalized version of the same log-normal shape~\cite{PiSasaki:2020lognormal,LVK:2025O4a},
\begin{equation}
 \mathcal P_\zeta(k)=\frac{A_\zeta}{\sqrt{2\pi}\Delta}
 \exp\!\left[-\frac{\ln^2(k/k_*)}{2\Delta^2}\right],
 \qquad \int d\ln k\,\mathcal P_\zeta=A_\zeta.
 \label{eq:lvk_lognormal}
\end{equation}
\begin{bluechange}
For $a_0=1$, the present observed frequency corresponding to a comoving
wavenumber is
\begin{equation}
 f_0=\frac{c k}{2\pi}
 \simeq 1.55\times10^{-15}
 \left(\frac{k}{\mathrm{Mpc}^{-1}}\right)\mathrm{Hz},
 \label{eq:k_to_frequency}
\end{equation}
which makes the connection between the peak scale $k_*$ and a detector band
explicit.
\end{bluechange}
Representative $95\%$ upper limits quoted by that analysis are \cite{LVK:2025O4a}
\begin{center}
\begin{tabular}{c|ccc}
\toprule
 & $k_*=10^{15}\,{\rm Mpc}^{-1}$ & $10^{17}\,{\rm Mpc}^{-1}$ & $10^{19}\,{\rm Mpc}^{-1}$\\
\midrule
$\Delta\rightarrow0$ & 1.44 & 0.01 & 0.15\\
$\Delta=1$ & 0.96 & 0.05 & 2.12\\
\bottomrule
\end{tabular}
\end{center}
These constrain the two-point normalization, not a connected four-point amplitude. Entries with $A_\zeta=\mathcal O(1)$ lie outside the conservative perturbative regime of the Gaussian second-order SIGW calculation and are quoted only as likelihood limits; the benchmark used below has a peak scalar power well below unity.

\subsection{Optimal statistic and idealized sampling estimate}
For a connected trispectrum template $\mathcal T_h$, an optimal weak-signal statistic has the schematic network form \cite{Ciprini:2026pvz}
\begin{equation}
 {\rm SNR}_4^2=T\sum_{abcd}\int df_1df_2df_3\,
 \frac{|\Gamma_{abcd}^{(4)}(f_1,f_2,f_3)\mathcal T_h(f_1,f_2,f_3)|^2}
 {\mathcal N_{abcd}(f_1,f_2,f_3)},
 \label{eq:optimal_fourpoint_snr}
\end{equation}
where frequency conservation fixes the fourth frequency, $\Gamma_{abcd}^{(4)}$ is the \black{four-channel network response}, and $\mathcal N_{abcd}$ includes instrumental, disconnected Gaussian, correlated-noise, and foreground terms. Equation~\eqref{eq:optimal_fourpoint_snr}, with a specified normalization convention, is required for a detector-specific prediction.

For orientation only, consider independent band-power samples in a narrow band and define
\begin{equation}
 r_s=\frac{S_h}{S_h+N_h+F_h},
 \qquad N_{\rm eff}=\xi_T T\Delta f,
 \label{eq:signal_fraction}
\end{equation}
where the convention-dependent factor $\xi_T=\mathcal O(1)$ counts independent real or complex Fourier samples. If the normalized estimator has variance coefficient $C_4$, then
\begin{equation}
 \sigma_\kappa\simeq\sqrt{\frac{C_4}{N_{\rm eff}}}\,r_s^{-2},
 \qquad
 \kappa_{n\sigma}\simeq n\sqrt{\frac{C_4}{\xi_TT\Delta f}}\,r_s^{-2}.
 \label{eq:kappa_threshold}
\end{equation}
The numerical illustration sets $\xi_T=1$ and $C_4=3$. Neither choice is universal. With those choices, four LISA-like years and $\Delta f=3$ mHz give $\kappa_{5\sigma}=1.41\times10^{-2}r_s^{-2}$, while one year and $\Delta f=100$ Hz give $1.54\times10^{-4}r_s^{-2}$. These are sample-count scalings, not LISA or 3G collaboration sensitivity curves.

If a chosen input template has amplitude $\kappa_{\rm in}$ and obeys the conditional damping law in Eq.~\eqref{eq:decoherence_kappa_scaling}, the same idealized benchmark yields
\begin{equation}
 D_k<D_k^{\rm max}=\frac14\ln\left(\frac{\kappa_{\rm in}}{\kappa_{n\sigma}}\right),
 \qquad \kappa_{\rm in}>\kappa_{n\sigma}.
 \label{eq:Dmax}
\end{equation}
Here $\kappa_{\rm in}=1$ is only a reference normalization; it is not a universal maximum.

\begin{table}[!htbp]
\centering
\caption{Idealized sample-count benchmarks obtained from Eq.~\eqref{eq:kappa_threshold} for $r_s=1$, $\xi_T=1$, and $C_4=3$. These numbers quantify the statistical scale of a fourth-order measurement; foreground dilution increases the threshold by $r_s^{-2}$.}
\label{tab:fourpoint_benchmarks}
\begin{tabular}{lccc}
\toprule
Benchmark & $T$ & $\Delta f$ & $\kappa_{5\sigma}$ \\
\midrule
LISA-like narrow band & $4\,{\rm yr}$ & $3\,{\rm mHz}$ & $1.41\times10^{-2}$ \\
3G ground-based band & $1\,{\rm yr}$ & $100\,{\rm Hz}$ & $1.54\times10^{-4}$ \\
\bottomrule
\end{tabular}
\end{table}

\black{Figure~\ref{fig:fisher} provides a compact visualization of how a phenomenological coherence amplitude scales with the strength of the stochastic background. It does not replace Eq.~\eqref{eq:optimal_fourpoint_snr}; rather, it illustrates the parameter-estimation logic and the rapid gain in sensitivity as the underlying SIGW signal becomes better measured.}

\begin{figure}[!htbp]
 \centering
 \includegraphics[width=0.90\textwidth]{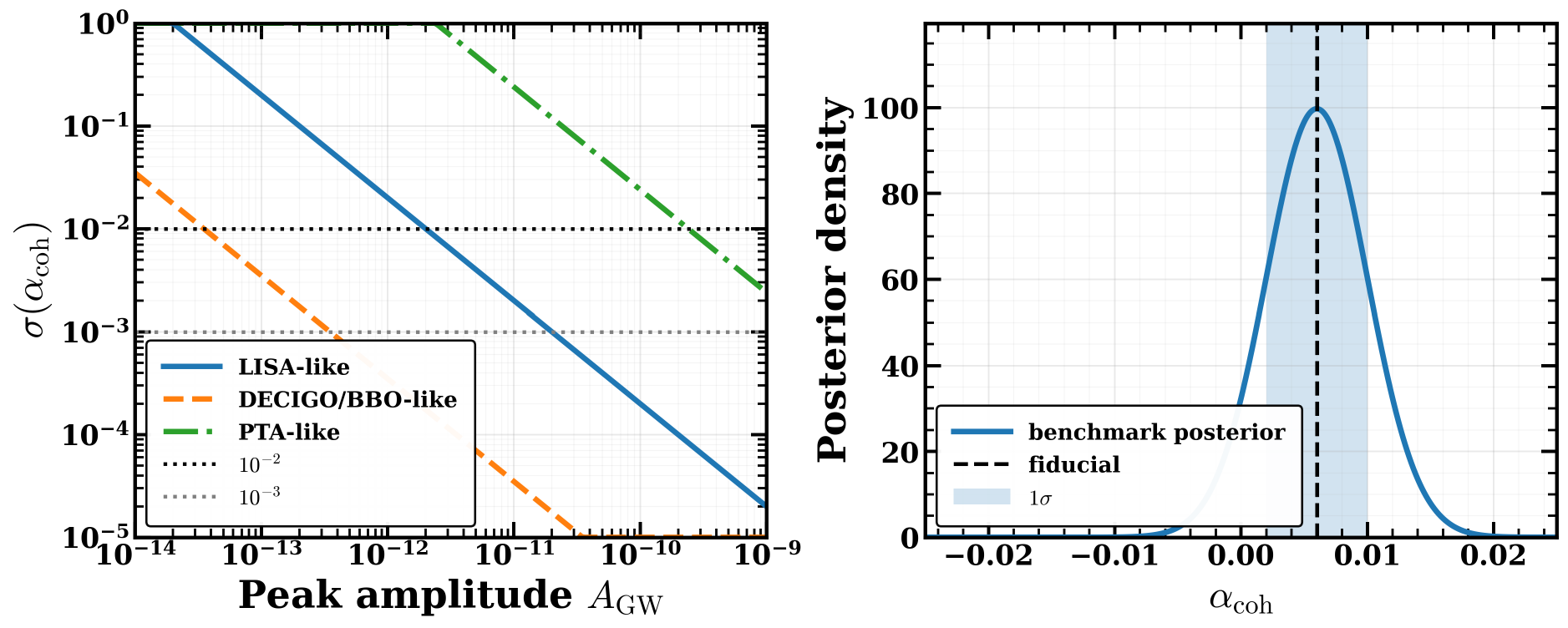}
 \caption{Illustrative Fisher sensitivity to a phenomenological coherence amplitude. The left panel shows the expected uncertainty as a function of the peak stochastic-background amplitude for representative detector bands, and the right panel shows a benchmark posterior. \black{The figure is an order-of-magnitude parameter-estimation illustration.} A detector-specific four-point forecast must instead use the response and covariance structure in Eq.~\eqref{eq:optimal_fourpoint_snr}, including foreground and correlated-noise nuisance terms.}
 \label{fig:fisher}
\end{figure}

Astrophysical populations and correlated instrumental or environmental noise are especially important for a fourth-order statistic because they can possess their own connected cumulants. They must therefore be included in $\mathcal N_{abcd}$ and, in a likelihood analysis, supplied with independent two-point and four-point nuisance templates. \black{For ground-based interferometers, correlated magnetic noise is a concrete example of an environmental contribution that must be controlled \cite{Thrane:2013npa}.}
\end{revision}

\section{Numerical consistency of the benchmark spectrum}
\label{sec:numerical_consistency}
\begin{revision}
For the ordinary two-point sector we use the standard radiation-era convolution
\begin{align}
 \Omega_{\rm GW,c}(k)&=\frac1{12}\int_0^\infty dv
 \int_{|1-v|}^{1+v}du
 \left[\frac{4v^2-(1+v^2-u^2)^2}{4uv}\right]^2
 \overline{I_{\rm RD}^2(u,v)}
 \mathcal P_\zeta(ku)\mathcal P_\zeta(kv),
 \label{eq:rd_uv_integral}\\
 \Omega_{\rm GW,0}(k)&=\Omega_{r,0}
 \frac{g_{*,\rho,c}}{g_{*,\rho,0}}
 \left(\frac{g_{*,s,0}}{g_{*,s,c}}\right)^{4/3}
 \Omega_{\rm GW,c}(k).
 \label{eq:rd_redshift_today}
\end{align}
The numerical implementation keeps the exact triangular $u$--$v$ domain and the oscillation-averaged radiation-era kernel. \black{Here $\overline{I_{\rm RD}^2}$ denotes the standard late-time, oscillation-averaged radiation-era kernel in the normalization of Eq.~\eqref{eq:rd_uv_integral}; in particular, the late-time rescaling that removes the explicit subhorizon $x^{-2}$ decay is included in this symbol. The formula therefore applies to modes whose relevant scalar sources re-enter during radiation domination \cite{Kohri:2018awv,Inomata:2018epa,Domenech:2021ztg}.} This calculation is used only to normalize the ordinary SIGW spectrum shown in Fig.~\ref{fig:kernel_signals}; it does not generate the connected four-point curve by itself.

The connected tensor statistic requires the scalar eight-point function because each tensor mode is quadratic in $\zeta$.  Accordingly, the exact prediction has the schematic decomposition
\begin{equation}
 \mathcal T_h[G^>_{\rm coh}]
 =\mathcal T_h[G^>_{\rm ref}]+\Delta\mathcal T_h^{\rm coh},
 \qquad
 \Delta\mathcal T_h^{\rm coh}\rightarrow0
 \quad(G^>_{\rm coh}\rightarrow G^>_{\rm ref}),
 \label{eq:trispectrum_baseline_split}
\end{equation}
and the corresponding measured residual is Eq.~\eqref{eq:obs_coherence_excess}.  The lower panel of Fig.~\ref{fig:kernel_signals} is therefore used as a phenomenological shape template for this residual, not as a substitute for the complete eight-point convolution.

This separation is essential for interpretation: ordinary SIGWs generated by Gaussian scalar perturbations are already non-Gaussian because of the quadratic source, so a nonzero four-point function alone is not a quantum signature. \black{The importance of connected and disconnected higher-order source contributions in SIGW calculations has been emphasized explicitly in Refs.~\cite{Adshead:2021ng,Perna:2024ng}.}  The proposed test is the joint consistency of the ordinary spectrum, the baseline-subtracted higher-order statistic, and, where a phase reference exists, the phase-sensitive correlator with one scalar state and one decoherence model.
\end{revision}

\section{Conclusions}
\label{sec:conclusions}
\begin{revision}
We have formulated residual quantum correlations in scalar-induced gravitational waves in a way that cleanly separates basis-dependent state parameters from observable correlation functions.  The primordial Gaussian state is described by $N_k^\zeta$ and $M_k^\zeta$ in a specified oscillator basis, while the actual scalar-to-tensor transfer is expressed through the full unequal-time Wightman function.  This prevents double counting of squeezing and removes the need to introduce an independent equal-time ``coherence power spectrum'' for a real scalar field.

The induced tensor field is quadratic in the scalar perturbation and is therefore generally non-Gaussian even when the scalar state is Gaussian.  Its second moments define an effective Gaussian reference state characterized by $d_k$ and $\gamma_k$, with the physicality condition $|\gamma_k|^2\le d_k(d_k+1)$.  The associated Gaussian discord is a useful covariance diagnostic, but the exact tensor occupation covariance contains an additional connected fourth-order cumulant.  For this reason, the relevant higher-order observable is not an un-subtracted connected four-point function, but the residual
\begin{equation}
 \Delta\kappa_{\rm coh}(k)
 =\kappa_N[G^>_{\rm coh}](k)-\kappa_N[G^>_{\rm ref}](k),
\end{equation}
where the reference model is constrained to reproduce the same diagonal two-point information.

\black{The figures illustrate complementary parts of this structure.}  The covariance plots show that entanglement can disappear while Gaussian discord and anomalous correlations remain.  The transfer plot shows how these surviving correlations can feed the tensor second moments within the stated Gaussian-reference approximation.  The combined $\Omega_{\rm GW}(f)$--$\kappa(f)$ figure emphasizes that the power spectrum and higher-order covariance encode different information, while the complex-plane figure makes explicit that the phase of $\gamma_k$ is meaningful only with a specified phase reference.  Finally, the Fisher benchmark and the general four-point signal-to-noise expression show why a connected-covariance search is substantially more demanding than an ordinary stochastic-background measurement.

For a phase-damping channel satisfying $M_k^\zeta\propto e^{-D_k}$, the leading Gaussian-reference contribution obeys the conditional hierarchy $|\gamma_k|\propto e^{-2D_k}$ and $\kappa_N^{\rm G}\propto e^{-4D_k}$.  These relations provide a direct way to translate a higher-order sensitivity into a constraint on residual coherence once a microscopic decoherence model specifies $D_k$.  They are not assumed to be universal: a general open-system evolution can also change $N_k^\zeta$, the phase of $M_k^\zeta$, and the source kernel.

The observational claim of this work is therefore deliberately specific.  A nonzero four-point function, Gaussian discord, or phase-sensitive correlator is not by itself proof of primordial quantumness, because classical nonstationary sources and astrophysical foregrounds can generate higher-order correlations.  The discriminating target is instead a consistent pattern across the ordinary SIGW spectrum, the baseline-subtracted connected statistic, and any phase-referenced correlator, all generated by one scalar state and one scalar-to-tensor transfer kernel.  This provides a concrete framework for confronting residual primordial coherence with future stochastic-background data while keeping the exact non-Gaussian SIGW baseline and detector response conceptually distinct.
\end{revision}

\appendix
\section{Unequal-time kernel and out-mode projection}
\label{app:kernels}
\begin{revision}
Writing
\begin{equation}
 \hat h_{\mathbf k,{\rm ind}}^\lambda(\tau_o)=
 \int^{\tau_o}d\tau\int\frac{d^3p}{(2\pi)^3}
 \mathcal I_\lambda(\mathbf k,\mathbf p;\tau_o,\tau)
 \hat\zeta_{\mathbf p}(\tau)\hat\zeta_{\mathbf k-\mathbf p}(\tau),
 \label{eq:app_h_source}
\end{equation}
the connected tensor two-point function for a Gaussian scalar state is exactly the two cross-contractions displayed in Eq.~\eqref{eq:tensor_unequal_time_transfer}. The third Wick pairing is a product of one-point source expectations and is removed by the connected prescription for nonzero external momentum.

For a normalized out-mode, the canonical projection is
\begin{equation}
 \hat b_{\mathbf k}^\lambda=i\left[
 v_k^{{\rm out}*}\hat\pi_{\mathbf k}^\lambda
 -v_k^{{\rm out}*\prime}\hat v_{\mathbf k}^\lambda
 \right]_{\tau=\tau_o},
 \label{eq:app_KG_projection}
\end{equation}
with the Wronskian in Eq.~\eqref{eq:tensor_out_expansion}. This projection must be applied to the full canonical tensor field. Applying it only to the retarded particular solution does not by itself define an operator with canonical commutator.
\end{revision}

\section{Two distinct Gaussian decoherence parameterizations}
\label{app:decoherence}
\begin{revision}
The additive covariance channel used in the diagnostic plots is
\begin{equation}
 \sigma_\zeta=\sigma_{\rm sq}+\epsilon_{\rm dec}\mathbb I_4.
 \label{eq:app_additive_noise}
\end{equation}
It leaves the off-diagonal covariance $c_k$ fixed and reduces $c_k/a_k$ by increasing the local variance. Reduced-density-matrix and open-system treatments of cosmological decoherence provide the microscopic setting in which such effective channels should ultimately be derived \black{\cite{Calzetta:1995ys,Kiefer:2006je,Burgess:2014eoa,Burgess:2022nwu}}. A phase-damping rate instead obeys Eq.~\eqref{eq:integrated_decoherence} and directly suppresses $M_k^\zeta$. These channels are not related by a universal map $\epsilon_{\rm dec}\leftrightarrow D_k$. A microscopic open-system calculation must provide the simultaneous evolution of $N_k^\zeta$, $M_k^\zeta$, and their phases.
\end{revision}

\section{General occupation covariance}
\label{app:kappa}
\begin{revision}
For two bosonic modes $a=\hat b_{\mathbf k}$ and $b=\hat b_{-\mathbf k}$, define $d_a=\langle a^\dagger a\rangle$, $d_b=\langle b^\dagger b\rangle$, $\gamma=\langle ab\rangle$, and $\eta=\langle a^\dagger b\rangle$. The fourth moment decomposes as
\begin{equation}
 \langle a^\dagger a b^\dagger b\rangle
 =d_ad_b+|\gamma|^2+|\eta|^2
 +\langle a^\dagger a b^\dagger b\rangle_c.
 \label{eq:app_fourth_decomposition}
\end{equation}
This identity defines the connected cumulant. For a Gaussian state the final term vanishes by Wick's theorem. For the exact SIGW state it generally does not vanish, even if the scalar source is Gaussian.
\end{revision}

\bibliographystyle{JHEP}
\bibliography{Bibliography}

@article{Guth:1980zm,
	author = "Guth, Alan H.",
	editor = "Fang, Li-Zhi and Ruffini, R.",
	title = "{The Inflationary Universe: A Possible Solution to the Horizon and Flatness Problems}",
	reportNumber = "SLAC-PUB-2576",
	doi = "10.1103/PhysRevD.23.347",
	journal = "Phys. Rev. D",
	volume = "23",
	pages = "347--356",
	year = "1981"
}

@article{Linde:1981mu,
	author = "Linde, Andrei D.",
	editor = "Fang, Li-Zhi and Ruffini, R.",
	title = "{A New Inflationary Universe Scenario: A Possible Solution of the Horizon, Flatness, Homogeneity, Isotropy and Primordial Monopole Problems}",
	reportNumber = "LEBEDEV-81-229",
	doi = "10.1016/0370-2693(82)91219-9",
	journal = "Phys. Lett. B",
	volume = "108",
	pages = "389--393",
	year = "1982"
}

@article{Albrecht:1982wi,
	author = "Albrecht, Andreas and Steinhardt, Paul J.",
	editor = "Fang, Li-Zhi and Ruffini, R.",
	title = "{Cosmology for Grand Unified Theories with Radiatively Induced Symmetry Breaking}",
	reportNumber = "UPR-0185T",
	doi = "10.1103/PhysRevLett.48.1220",
	journal = "Phys. Rev. Lett.",
	volume = "48",
	pages = "1220--1223",
	year = "1982"
}

@article{Mukhanov:1981xt,
	author = "Mukhanov, Viatcheslav F. and Chibisov, G. V.",
	title = "{Quantum Fluctuations and a Nonsingular Universe}",
	journal = "JETP Lett.",
	volume = "33",
	pages = "532--535",
	year = "1981"
}

@article{Mukhanov:1990me,
	author = "Mukhanov, Viatcheslav F. and Feldman, H. A. and Brandenberger, Robert H.",
	title = "{Theory of cosmological perturbations. Part 1. Classical perturbations. Part 2. Quantum theory of perturbations. Part 3. Extensions}",
	reportNumber = "BROWN-HET-796, BROWN-HET-800, BROWN-HET-780",
	doi = "10.1016/0370-1573(92)90044-Z",
	journal = "Phys. Rept.",
	volume = "215",
	pages = "203--333",
	year = "1992"
}

@article{Polarski:1995jg,
	author = "Polarski, David and Starobinsky, Alexei A.",
	title = "{Semiclassicality and decoherence of cosmological perturbations}",
	eprint = "gr-qc/9504030",
	archivePrefix = "arXiv",
	reportNumber = "LMPM-95-4",
	doi = "10.1088/0264-9381/13/3/006",
	journal = "Class. Quant. Grav.",
	volume = "13",
	pages = "377--392",
	year = "1996"
}

@article{Kiefer:1998qe,
	author = "Kiefer, Claus and Polarski, David and Starobinsky, Alexei A.",
	title = "{Quantum to classical transition for fluctuations in the early universe}",
	eprint = "gr-qc/9802003",
	archivePrefix = "arXiv",
	reportNumber = "THEP-97-33, FREIBURG-THEP-97-33, Freiburg THEP-97/33",
	doi = "10.1142/S0218271898000292",
	journal = "Int. J. Mod. Phys. D",
	volume = "7",
	pages = "455--462",
	year = "1998"
}

@article{Albrecht:1992kf,
	author = "Albrecht, Andreas and Ferreira, Pedro and Joyce, Michael and Prokopec, Tomislav",
	title = "{Inflation and squeezed quantum states}",
	eprint = "astro-ph/9303001",
	archivePrefix = "arXiv",
	reportNumber = "IMPERIAL-TP-92-93-21",
	doi = "10.1103/PhysRevD.50.4807",
	journal = "Phys. Rev. D",
	volume = "50",
	pages = "4807--4820",
	year = "1994"
}

@article{Grishchuk:1990bj,
	author = "Grishchuk, L. P. and Sidorov, Yu. V.",
	title = "{Squeezed quantum states of relic gravitons and primordial density fluctuations}",
	doi = "10.1103/PhysRevD.42.3413",
	journal = "Phys. Rev. D",
	volume = "42",
	pages = "3413--3421",
	year = "1990"
}

@article{Schlosshauer:2014pgr,
	author = "Schlosshauer, Maximilian",
	title = "{The quantum-to-classical transition and decoherence}",
	eprint = "1404.2635",
	archivePrefix = "arXiv",
	primaryClass = "quant-ph",
	month = "4",
	year = "2014"
}

@article{Zurek:2003zz,
	author = "Zurek, Wojciech Hubert",
	title = "{Decoherence, einselection, and the quantum origins of the classical}",
	eprint = "quant-ph/0105127",
	archivePrefix = "arXiv",
	doi = "10.1103/RevModPhys.75.715",
	journal = "Rev. Mod. Phys.",
	volume = "75",
	pages = "715--775",
	year = "2003"
}

@article{Kiefer:2006je,
	author = "Kiefer, Claus and Lohmar, Ingo and Polarski, David and Starobinsky, Alexei A.",
	title = "{Pointer states for primordial fluctuations in inflationary cosmology}",
	eprint = "astro-ph/0610700",
	archivePrefix = "arXiv",
	doi = "10.1088/0264-9381/24/7/002",
	journal = "Class. Quant. Grav.",
	volume = "24",
	pages = "1699--1718",
	year = "2007"
}

@article{Modi:2012baj,
	author = "Modi, Kavan and Brodutch, Aharon and Cable, Hugo and Paterek, Tomasz and Vedral, Vlatko",
	title = "{The classical-quantum boundary for correlations: Discord and related measures}",
	eprint = "1112.6238",
	archivePrefix = "arXiv",
	primaryClass = "quant-ph",
	doi = "10.1103/RevModPhys.84.1655",
	journal = "Rev. Mod. Phys.",
	volume = "84",
	number = "4",
	pages = "1655",
	year = "2012"
}

@article{Adesso:2014npz,
	author = "Adesso, Gerardo and Ragy, Sammy and Lee, Antony R.",
	title = "{Continuous Variable Quantum Information: Gaussian States and Beyond}",
	eprint = "1401.4679",
	archivePrefix = "arXiv",
	primaryClass = "quant-ph",
	doi = "10.1142/s1230161214400010",
	journal = "Open Syst. Info. Dyn.",
	volume = "21",
	number = "01n02",
	pages = "1440001",
	year = "2014"
}

@article{Weedbrook:2011wxo,
	author = "Weedbrook, Christian and Pirandola, Stefano and Garc{\'\i}a-Patr{\'o}n, Ra{\'u}l and Cerf, Nicolas J. and Ralph, Timothy C. and Shapiro, Jeffrey H. and Lloyd, Seth",
	title = "{Gaussian quantum information}",
	eprint = "1110.3234",
	archivePrefix = "arXiv",
	primaryClass = "quant-ph",
	doi = "10.1103/RevModPhys.84.621",
	journal = "Rev. Mod. Phys.",
	volume = "84",
	number = "2",
	pages = "621",
	year = "2012"
}

@article{Giorda:2010wpy,
	author = "Giorda, Paolo and Paris, Matteo G. A.",
	title = "{Gaussian Quantum Discord}",
	eprint = "1003.3207",
	archivePrefix = "arXiv",
	primaryClass = "quant-ph",
	doi = "10.1103/PhysRevLett.105.020503",
	journal = "Phys. Rev. Lett.",
	volume = "105",
	number = "2",
	pages = "020503",
	year = "2010"
}

@article{Martin:2015qta,
	author = "Martin, Jerome and Vennin, Vincent",
	title = "{Quantum Discord of Cosmic Inflation: Can we Show that CMB Anisotropies are of Quantum-Mechanical Origin?}",
	eprint = "1510.04038",
	archivePrefix = "arXiv",
	primaryClass = "astro-ph.CO",
	doi = "10.1103/PhysRevD.93.023505",
	journal = "Phys. Rev. D",
	volume = "93",
	number = "2",
	pages = "023505",
	year = "2016"
}

@article{Domenech:2021ztg,
	author = "Dom{\`e}nech, Guillem",
	title = "{Scalar Induced Gravitational Waves Review}",
	eprint = "2109.01398",
	archivePrefix = "arXiv",
	primaryClass = "gr-qc",
	doi = "10.3390/universe7110398",
	journal = "Universe",
	volume = "7",
	number = "11",
	pages = "398",
	year = "2021"
}

@article{Ananda:2006af,
	author = "Ananda, Kishore N. and Clarkson, Chris and Wands, David",
	title = "{The Cosmological gravitational wave background from primordial density perturbations}",
	eprint = "gr-qc/0612013",
	archivePrefix = "arXiv",
	doi = "10.1103/PhysRevD.75.123518",
	journal = "Phys. Rev. D",
	volume = "75",
	pages = "123518",
	year = "2007"
}

@article{Saito:2008jc,
	author = "Saito, Ryo and Yokoyama, Jun'ichi",
	title = "{Gravitational wave background as a probe of the primordial black hole abundance}",
	eprint = "0812.4339",
	archivePrefix = "arXiv",
	primaryClass = "astro-ph",
	reportNumber = "RESCEU-63-08",
	doi = "10.1103/PhysRevLett.102.161101",
	journal = "Phys. Rev. Lett.",
	volume = "102",
	pages = "161101",
	year = "2009",
	note = "[Erratum: Phys.Rev.Lett. 107, 069901 (2011)]"
}

@article{Baumann:2007zm,
	author = "Baumann, Daniel and Steinhardt, Paul J. and Takahashi, Keitaro and Ichiki, Kiyotomo",
	title = "{Gravitational Wave Spectrum Induced by Primordial Scalar Perturbations}",
	eprint = "hep-th/0703290",
	archivePrefix = "arXiv",
	doi = "10.1103/PhysRevD.76.084019",
	journal = "Phys. Rev. D",
	volume = "76",
	pages = "084019",
	year = "2007"
}

@article{Carr:2021bzv,
	author = "Carr, Bernard and Kuhnel, Florian",
	title = "{Primordial black holes as dark matter candidates}",
	eprint = "2110.02821",
	archivePrefix = "arXiv",
	primaryClass = "astro-ph.CO",
	doi = "10.21468/SciPostPhysLectNotes.48",
	journal = "SciPost Phys. Lect. Notes",
	volume = "48",
	pages = "1",
	year = "2022"
}

@article{LISA:2017pwj,
	author = "Amaro-Seoane, Pau and others",
	collaboration = "LISA",
	title = "{Laser Interferometer Space Antenna}",
	eprint = "1702.00786",
	archivePrefix = "arXiv",
	primaryClass = "astro-ph.IM",
	month = "2",
	year = "2017"
}

@article{NANOGrav:2023gor,
	author = "Agazie, Gabriella and others",
	collaboration = "NANOGrav",
	title = "{The NANOGrav 15 yr Data Set: Evidence for a Gravitational-wave Background}",
	eprint = "2306.16213",
	archivePrefix = "arXiv",
	primaryClass = "astro-ph.HE",
	doi = "10.3847/2041-8213/acdac6",
	journal = "Astrophys. J. Lett.",
	volume = "951",
	number = "1",
	pages = "L8",
	year = "2023"
}

@article{EPTA:2023fyk,
	author = "Antoniadis, J. and others",
	collaboration = "EPTA, InPTA:",
	title = "{The second data release from the European Pulsar Timing Array - III. Search for gravitational wave signals}",
	eprint = "2306.16214",
	archivePrefix = "arXiv",
	primaryClass = "astro-ph.HE",
	doi = "10.1051/0004-6361/202346844",
	journal = "Astron. Astrophys.",
	volume = "678",
	pages = "A50",
	year = "2023"
}

@article{Xu:2023wog,
	author = "Xu, Heng and others",
	title = "{Searching for the Nano-Hertz Stochastic Gravitational Wave Background with the Chinese Pulsar Timing Array Data Release I}",
	eprint = "2306.16216",
	archivePrefix = "arXiv",
	primaryClass = "astro-ph.HE",
	doi = "10.1088/1674-4527/acdfa5",
	journal = "Res. Astron. Astrophys.",
	volume = "23",
	number = "7",
	pages = "075024",
	year = "2023"
}

@article{Adesso:2010bfw,
	author = "Adesso, Gerardo and Datta, Animesh",
	title = "{Quantum versus Classical Correlations in Gaussian States}",
	eprint = "1003.4979",
	archivePrefix = "arXiv",
	primaryClass = "quant-ph",
	doi = "10.1103/PhysRevLett.105.030501",
	journal = "Phys. Rev. Lett.",
	volume = "105",
	number = "3",
	pages = "030501",
	year = "2010"
}

@article{Duan:2000awi,
	author = "Duan, Lu-Ming and Giedke, G. and Cirac, J. I. and Zoller, P.",
	title = "{Inseparability Criterion for Continuous Variable Systems}",
	eprint = "quant-ph/9908056",
	archivePrefix = "arXiv",
	doi = "10.1103/PhysRevLett.84.2722",
	journal = "Phys. Rev. Lett.",
	volume = "84",
	number = "12",
	pages = "2722",
	year = "2000"
}

@article{Henderson:2001wrr,
	author = "Henderson, L. and Vedral, V.",
	title = "{Classical, quantum and total correlations}",
	eprint = "quant-ph/0105028",
	archivePrefix = "arXiv",
	doi = "10.1088/0305-4470/34/35/315",
	journal = "J. Phys. A",
	volume = "34",
	number = "35",
	pages = "6899",
	year = "2001"
}

@article{Burgess:2014eoa,
	author = "Burgess, C. P. and Holman, R. and Tasinato, G. and Williams, M.",
	title = "{EFT Beyond the Horizon: Stochastic Inflation and How Primordial Quantum Fluctuations Go Classical}",
	eprint = "1408.5002",
	archivePrefix = "arXiv",
	primaryClass = "hep-th",
	reportNumber = "CERN-PH-TH-2014-142",
	doi = "10.1007/JHEP03(2015)090",
	journal = "JHEP",
	volume = "03",
	pages = "090",
	year = "2015"
}

@article{Kawamura:2020pcg,
	author = "Kawamura, Seiji and others",
	title = "{Current status of space gravitational wave antenna DECIGO and B-DECIGO}",
	eprint = "2006.13545",
	archivePrefix = "arXiv",
	primaryClass = "gr-qc",
	doi = "10.1093/ptep/ptab019",
	journal = "PTEP",
	volume = "2021",
	number = "5",
	pages = "05A105",
	year = "2021"
}

@article{Kohri:2018awv,
	author = "Kohri, Kazunori and Terada, Takahiro",
	title = "{Semianalytic calculation of gravitational wave spectrum nonlinearly induced from primordial curvature perturbations}",
	eprint = "1804.08577",
	archivePrefix = "arXiv",
	primaryClass = "gr-qc",
	reportNumber = "KEK-TH-2046, KEK-COSMO-223",
	doi = "10.1103/PhysRevD.97.123532",
	journal = "Phys. Rev. D",
	volume = "97",
	number = "12",
	pages = "123532",
	year = "2018"
}

@article{Weinberg:2005vy,
    author = "Weinberg, Steven",
    title = "{Quantum contributions to cosmological correlations}",
    eprint = "hep-th/0506236",
    archivePrefix = "arXiv",
    reportNumber = "UTTG-01-05",
    doi = "10.1103/PhysRevD.72.043514",
    journal = "Phys. Rev. D",
    volume = "72",
    pages = "043514",
    year = "2005"
}

@article{Adshead:2009cb,
    author = "Adshead, Peter and Easther, Richard and Lim, Eugene A.",
    title = "{The 'in-in' Formalism and Cosmological Perturbations}",
    eprint = "0904.4207",
    archivePrefix = "arXiv",
    primaryClass = "hep-th",
    doi = "10.1103/PhysRevD.80.083521",
    journal = "Phys. Rev. D",
    volume = "80",
    pages = "083521",
    year = "2009"
}

@article{Kiefer:1998jk,
    author = "Kiefer, Claus and Polarski, David",
    title = "{Emergence of classicality for primordial fluctuations: Concepts and analogies}",
    eprint = "gr-qc/9805014",
    archivePrefix = "arXiv",
    reportNumber = "FREIBURG-THEP-98-7, FREIBURG-THP-98-7, Freiburg THP-98/7",
    doi = "10.1002/andp.2090070302",
    journal = "Annalen Phys.",
    volume = "7",
    pages = "137--158",
    year = "1998"
}

@article{Vidal:2002zz,
    author = "Vidal, G. and Werner, R. F.",
    title = "{Computable measure of entanglement}",
    eprint = "quant-ph/0102117",
    archivePrefix = "arXiv",
    doi = "10.1103/PhysRevA.65.032314",
    journal = "Phys. Rev. A",
    volume = "65",
    pages = "032314",
    year = "2002"
}

@article{Inomata:2018epa,
	author = "Inomata, Keisuke and Nakama, Tomohiro",
	title = "{Gravitational waves induced by scalar perturbations as probes of the small-scale primordial spectrum}",
	eprint = "1812.00674",
	archivePrefix = "arXiv",
	primaryClass = "astro-ph.CO",
	reportNumber = "IPMU 18-0200",
	doi = "10.1103/PhysRevD.99.043511",
	journal = "Phys. Rev. D",
	volume = "99",
	number = "4",
	pages = "043511",
	year = "2019"
}

@article{Calzetta:1995ys,
	author = "Calzetta, E. and Hu, B. L.",
	title = "{Quantum fluctuations, decoherence of the mean field, and structure formation in the early universe}",
	eprint = "gr-qc/9505046",
	archivePrefix = "arXiv",
	reportNumber = "IASSNS-HEP-95-38, UMD-PP-95-083",
	doi = "10.1103/PhysRevD.52.6770",
	journal = "Phys. Rev. D",
	volume = "52",
	pages = "6770--6788",
	year = "1995"
}

@article{Burgess:2022nwu,
	author = "Burgess, C. P. and Holman, R. and Kaplanek, Greg and Martin, Jerome and Vennin, Vincent",
	title = "{Minimal decoherence from inflation}",
	eprint = "2211.11046",
	archivePrefix = "arXiv",
	primaryClass = "hep-th",
	reportNumber = "CERN-TH-2022-174; Imperial/TP/2022/GK/02",
	doi = "10.1088/1475-7516/2023/07/022",
	journal = "JCAP",
	volume = "07",
	pages = "022",
	year = "2023"
}

@article{Mukherjee:2025dcv,
	author = "Mukherjee, Dipayan and Ragavendra, H. V. and Sethi, Shiv K.",
	title = "{Scalar-induced gravitational waves from coherent initial states}",
	eprint = "2506.23798",
	archivePrefix = "arXiv",
	primaryClass = "astro-ph.CO",
	doi = "10.1103/qd1s-9fxl",
	journal = "Phys. Rev. D",
	volume = "113",
	number = "2",
	pages = "023533",
	year = "2026"
}

@article{Isar:2013vvd,
	author = "Isar, Aurelian",
	title = "{Quantum correlations of two-mode Gaussian systems in a thermal environment}",
	eprint = "1301.0549",
	archivePrefix = "arXiv",
	primaryClass = "quant-ph",
	doi = "10.1088/0031-8949/2013/T153/014035",
	journal = "Phys. Scripta",
	volume = "T153",
	pages = "014035",
	year = "2013"
}

@article{Thrane:2013npa,
	author = "Thrane, Eric and Christensen, Nelson and Schofield, Robert",
	title = "{Correlated magnetic noise in global networks of gravitational-wave interferometers: observations and implications}",
	eprint = "1303.2613",
	archivePrefix = "arXiv",
	primaryClass = "astro-ph.IM",
	doi = "10.1103/PhysRevD.87.123009",
	journal = "Phys. Rev. D",
	volume = "87",
	pages = "123009",
	year = "2013"
}

@article{Ciprini:2026pvz,
	author = "Ciprini, Martina and Marcelli, Maria Lucia and Tasinato, Gianmassimo",
	title = "{Probing gravitational-wave four-point correlators}",
	eprint = "2603.15514",
	archivePrefix = "arXiv",
	primaryClass = "astro-ph.CO",
	doi = "10.1103/3pd3-r3x8",
	journal = "Phys. Rev. D",
	volume = "113",
	number = "10",
	pages = "103544",
	year = "2026"
}

@article{Romano:2016dpx,
	author = "Romano, Joseph D. and Cornish, Neil J.",
	title = "{Detection methods for stochastic gravitational-wave backgrounds: a unified treatment}",
	eprint = "1608.06889",
	archivePrefix = "arXiv",
	primaryClass = "gr-qc",
	doi = "10.1007/s41114-017-0004-1",
	journal = "Living Rev. Rel.",
	volume = "20",
	number = "1",
	pages = "2",
	year = "2017"
}

@article{Caprini:2018mtu,
	author = "Caprini, Chiara and Figueroa, Daniel G.",
	title = "{Cosmological backgrounds of gravitational waves}",
	eprint = "1801.04268",
	archivePrefix = "arXiv",
	primaryClass = "astro-ph.CO",
	doi = "10.1088/1361-6382/aac608",
	journal = "Class. Quant. Grav.",
	volume = "35",
	number = "16",
	pages = "163001",
	year = "2018"
}

@article{Giovannini:2010xg,
	author = "Giovannini, Massimo",
	title = "{Hanbury Brown-Twiss interferometry and second-order correlations of inflaton quanta}",
	eprint = "1011.1673",
	archivePrefix = "arXiv",
	primaryClass = "astro-ph.CO",
	reportNumber = "CERN-PH-TH-2010-242",
	doi = "10.1103/PhysRevD.83.023515",
	journal = "Phys. Rev. D",
	volume = "83",
	pages = "023515",
	year = "2011"
}

@article{Toccacelo:2026hcz,
	author = "Toccacelo, Kristian and Beitel, Thomas and Andersen, Ulrik Lund and Pikovski, Igor",
	title = "{Quantum State Characterization of Gravitational Waves via Graviton Counting Statistics}",
	eprint = "2602.09125",
	archivePrefix = "arXiv",
	primaryClass = "quant-ph",
	month = "2",
	year = "2026"
}

@article{TianQin:2015yph,
	author = "Luo, Jun and others",
	collaboration = "TianQin",
	title = "{TianQin: a space-borne gravitational wave detector}",
	eprint = "1512.02076",
	archivePrefix = "arXiv",
	primaryClass = "astro-ph.IM",
	doi = "10.1088/0264-9381/33/3/035010",
	journal = "Class. Quant. Grav.",
	volume = "33",
	number = "3",
	pages = "035010",
	year = "2016"
}

@article{Crowder:2005nr,
	author = "Crowder, Jeff and Cornish, Neil J.",
	title = "{Beyond LISA: Exploring future gravitational wave missions}",
	eprint = "gr-qc/0506015",
	archivePrefix = "arXiv",
	doi = "10.1103/PhysRevD.72.083005",
	journal = "Phys. Rev. D",
	volume = "72",
	pages = "083005",
	year = "2005"
}

@article{Burke-Spolaor:2018bvk,
	author = "Burke-Spolaor, Sarah and others",
	title = "{The Astrophysics of Nanohertz Gravitational Waves}",
	eprint = "1811.08826",
	archivePrefix = "arXiv",
	primaryClass = "astro-ph.HE",
	doi = "10.1007/s00159-019-0115-7",
	journal = "Astron. Astrophys. Rev.",
	volume = "27",
	number = "1",
	pages = "5",
	year = "2019"
}

@article{PiSasaki:2020lognormal,
    author = "Pi, Shi and Sasaki, Misao",
    title = "{Gravitational Waves Induced by Scalar Perturbations with a Lognormal Peak}",
    eprint = "2005.12306",
    archivePrefix = "arXiv",
    primaryClass = "gr-qc",
    reportNumber = "YITP-20-75, IPMU20-0054",
    doi = "10.1088/1475-7516/2020/09/037",
    journal = "JCAP",
    volume = "09",
    pages = "037",
    year = "2020"
}

@article{LVK:2025O4a,
    author = "{LIGO Scientific Collaboration and Virgo Collaboration and KAGRA Collaboration}",
    title = "{Cosmological and High Energy Physics implications from gravitational-wave background searches in LIGO-Virgo-KAGRA's O1-O4a runs}",
    eprint = "2510.26848",
    archivePrefix = "arXiv",
    primaryClass = "gr-qc",
    year = "2025"
}

@article{Simon:2000zz,
    author = "Simon, R.",
    title = "{Peres--Horodecki Separability Criterion for Continuous Variable Systems}",
    eprint = "quant-ph/9909044",
    archivePrefix = "arXiv",
    doi = "10.1103/PhysRevLett.84.2726",
    journal = "Phys. Rev. Lett.",
    volume = "84",
    pages = "2726--2729",
    year = "2000"
}

@article{Thrane:2013ng,
    author = "Thrane, Eric",
    title = "{Measuring the non-Gaussian stochastic gravitational-wave background: a method for realistic interferometer data}",
    eprint = "1301.0263",
    archivePrefix = "arXiv",
    primaryClass = "astro-ph.IM",
    doi = "10.1103/PhysRevD.87.043009",
    journal = "Phys. Rev. D",
    volume = "87",
    number = "4",
    pages = "043009",
    year = "2013"
}

@article{Seto:2009ng,
    author = "Seto, Naoki",
    title = "{Non-Gaussianity analysis of a gravitational wave background made by short-duration burst signals}",
    doi = "10.1103/PhysRevD.80.043003",
    journal = "Phys. Rev. D",
    volume = "80",
    pages = "043003",
    year = "2009"
}

@article{Adshead:2021ng,
    author = "Adshead, Peter and Lozanov, Kaloian D. and Weiner, Zachary J.",
    title = "{Non-Gaussianity and the induced gravitational wave background}",
    eprint = "2105.01659",
    archivePrefix = "arXiv",
    primaryClass = "astro-ph.CO",
    doi = "10.1088/1475-7516/2021/10/080",
    journal = "JCAP",
    volume = "10",
    pages = "080",
    year = "2021"
}

@article{Perna:2024ng,
    author = "Perna, Gabriele and Testini, Chiara and Ricciardone, Angelo and Matarrese, Sabino",
    title = "{Fully non-Gaussian Scalar-Induced Gravitational Waves}",
    eprint = "2403.06962",
    archivePrefix = "arXiv",
    primaryClass = "gr-qc",
    doi = "10.1088/1475-7516/2024/05/086",
    journal = "JCAP",
    volume = "05",
    pages = "086",
    year = "2024"
}

\end{document}